\PassOptionsToPackage{table}{xcolor}
\documentclass[10pt,journal,compsoc]{IEEEtran}
\IEEEoverridecommandlockouts
\usepackage{longtable}
\usepackage[noadjust]{cite}
\usepackage{amsmath,amssymb,amsfonts}
\usepackage{algorithmic}
\usepackage{graphicx}

\usepackage{comment}
\usepackage{textcomp}
\usepackage{xcolor}
\usepackage{adjustbox}
\usepackage{textcomp}
\usepackage[numbered]{bookmark}
\usepackage{soul}
\usepackage{booktabs}
\usepackage{multirow}
\usepackage{float}
\usepackage[many]{tcolorbox}
\usepackage{graphicx}
\usepackage{url}
\usepackage{tabularx} 
\usepackage[justification=centering]{caption}
\usepackage{caption}
\usepackage{subcaption}
\usepackage{tablefootnote}
\usepackage[switch]{lineno}
\usepackage{ragged2e}
\usepackage{dirtytalk}
\usepackage{multicol}
\usepackage{csquotes}
\usepackage{pgfplots, pgfplotstable}
\usepackage{pgf-pie}
\usepackage{colortbl}
\usepackage{xspace}
\newcommand{\ie}{\textit{i.e., \xspace}}
\newcommand{\eg}{\textit{e.g., \xspace}}
\newcommand{\etal}{\textit{et al. \xspace}}
\usepackage{array}
\usepackage{balance}
\usepackage{supertabular} 
\usepackage{fontawesome}
\usepackage{listings}
\usepackage[table]{xcolor}
\newcommand{\initialpool}{1,367\xspace}
\newcommand{\finalpool}{83\xspace}
\newcommand{\noduplicate}{943\xspace}
\newcommand{\snowballing}{17\xspace}
\newcommand{\tool}{37\xspace}
\newcommand{\toolaccuracy}{11\xspace}
\newcommand{\toolonline}{18\xspace}
\newcommand{\metric}{14\xspace}
\newcommand{\titleabstract}{114\xspace}
\newcommand{\fulltext}{66\xspace}
\newcommand{\reference}{1,958\xspace}
\newcommand{\lstbg}[3][0pt]{{\fboxsep#1\colorbox{#2}{\strut #3}}}
\lstdefinelanguage{diff}{
  basicstyle=\ttfamily\scriptsize,
  morecomment=[f][\lstbg{red!20}]-,
  morecomment=[f][\lstbg{green!20}]+,
  morecomment=[f][\textit]{@@},
}

\newtcolorbox{boxK}{
    sharpish corners, 
    boxrule = 0pt,
    toprule = 4.5pt, 
    enhanced,
    fuzzy shadow = {0pt}{-2pt}{-0.5pt}{0.5pt}{black!35} 
}

\definecolor{javared}{rgb}{0.6,0,0} 
\definecolor{javagreen}{rgb}{0.25,0.5,0.35} 
\definecolor{javapurple}{rgb}{0.5,0,0.35} 
\definecolor{javadocblue}{rgb}{0.25,0.35,0.75} 
 
\pgfplotstableread[col sep=comma,header=true]{
Category,1,2,3
1998,0.5,0,0
1999, 0, 0, 0
2000, 0, 0, 0
2001,0,100,0
2002, 0, 0, 0
2003, 0, 0, 0
2004,33.33,66.66,0 
2005, 0, 0, 0
2006, 0, 0, 0
2007,0,100,0 
2008,0,50,50 
2009,33.33,66.66,0 
2010, 0, 0, 0
2011,33.33,33.33,33.33 
2012,33.33,33.33,33.33 
2013,0,50,50 
2014,0,100,0 
2015,0,100,0 
2016,0,50,50 
2017,0,90,10
2018,0,100,0 
2019,33.33,66.66,0 
2020,0,10,90
2021,0,50,50  
2022,0,80,20 
2023,0,50,50 
}\dataA

\pgfplotstablecreatecol[
 create col/expr={
    \thisrow{1} + \thisrow{2} + \thisrow{3} 
 }
]{sum}{\dataA}

\pgfplotsset{
  percentage plot/.style={
   point meta=explicit,
    every node near coord/.append style={
      font=\tiny,
      color=black,
    },
    nodes near coords={
      \pgfmathtruncatemacro\iszero{\originalvalue==0}
      \ifnum\iszero=0
      \pgfmathprintnumber[fixed,fixed zerofill,precision=1]{\pgfplotspointmeta}
      \fi
    },
    yticklabel=\pgfmathprintnumber{\tick}\,$\%$,
    ymin=0,
    ymax=100.01, 
    visualization depends on={y \as \originalvalue},
    enlarge x limits={abs=10mm}
  },
  percentage series/.style={
    table/x expr=\coordindex, 
    table/y expr=(\thisrow{#1}/\thisrow{sum}*100),
    table/meta=#1
    }
}

\newcommand{\eman}[1]{\textcolor{violet}{{\it [Eman says: #1]}}}

\newcommand{\ali}[1]{\textcolor{green}{{\it [Ali: #1]}}}

\usepackage[flushleft]{threeparttable}
\usepackage{supertabular}
\DeclareCaptionType{TextBox}
\newcolumntype{L}{>{\arraybackslash}m{16cm}}
\newcolumntype{C}[1]{>{\centering\let\newline\\arraybackslash\hspace{0pt}}m{#1}}
\newcolumntype{R}[1]{>{\raggedleft\let\newline\\arraybackslash\hspace{0pt}}m{#1}}
\def\BibTeX{{\rm B\kern-.05em{\sc i\kern-.025em b}\kern-.08em
    T\kern-.1667em\lower.7ex\hbox{E}\kern-.125emX}}
    
    \newcommand{\RQone}{\textit{What approaches were considered by the PSs to recommend Extract Method refactoring?\xspace}}
\newcommand{\RQtwo}{\textit{What are the main characteristics of Extract Method recommendation tools?\xspace}}
\newcommand{\RQthree}{\textit{What are the datasets, and benchmarks used
for evaluating and validating Extract Method recommendation approaches?\xspace}}
\begin{document}

\title{\huge Behind the Intent of Extract Method Refactoring 
\\ \large \textcolor{black}{A Systematic Literature Review}\\
}

\author{Eman Abdullah AlOmar,~\IEEEmembership{Member,~IEEE,}
        Mohamed Wiem Mkaouer,~\IEEEmembership{Member,~IEEE,}
        and~Ali~Ouni,~\IEEEmembership{Member,~IEEE}
\IEEEcompsocitemizethanks{\IEEEcompsocthanksitem EA. AlOmar is with the School of Systems and Enterprises, Stevens Institute of Technology, Hoboken,
NJ, 07030 USA.\protect\\
E-mail: ealomar@stevens.edu
\IEEEcompsocthanksitem MW. Mkaouer is with the College of Innovation and Technology, University of Michigan-Flint, Flint, MI 48502 USA.
E-mail: mmkaouer@umich.edu
\IEEEcompsocthanksitem A. Ouni is with the Department of Software Engineering and IT, ETS Montreal, University of Quebec, H3C 3P8 Montreal,
QC, Canada. E-mail: ali.ouni@etsmtl.ca}
\thanks{Manuscript received May 7, 2023.}}

\markboth{Transactions on Software Engineering}%
{AlOmar \MakeLowercase{\textit{et al.}}: Bare Advanced Demo of IEEEtran.cls for IEEE Computer Society Journals}
\IEEEtitleabstractindextext{%
\begin{abstract}
\textbf{Background:} 
Code refactoring is widely recognized as an essential software engineering practice to improve the understandability and maintainability of the source code. The \textit{Extract Method} refactoring is considered as \say{Swiss army knife} of refactorings, as developers often apply it to improve their code quality, \eg decompose long code fragments, reduce code complexity, eliminate duplicated code, etc. In recent years, several studies attempted to recommend \textit{Extract Method} refactorings 
 allowing the collection, analysis, and revelation of actionable data-driven insights about refactoring practices within software projects. 

\textbf{Aim:} 
\textcolor{black}{In this paper, we aim at reviewing the current body of knowledge} on existing Extract Method refactoring research and explore their limitations and potential improvement opportunities for future research efforts. That is, \textit{Extract Method} is considered one of the most widely-used refactorings, but difficult to apply in practice as it involves low-level code changes such as statements, variables, parameters, return types, etc. Hence, researchers and practitioners begin to be aware of the state-of-the-art and identify new research opportunities in this context. 

\textbf{Method:} 
We review the body of knowledge related to \textit{Extract Method} refactoring in the form of a systematic literature review (SLR). After compiling an initial pool of \initialpool papers, we conducted a systematic selection and our final pool included \finalpool \textcolor{black}{primary studies}. We define \textcolor{black}{three} sets of research questions and systematically develop and refine a classification schema based on several criteria including their methodology, applicability, and degree of automation. 

\textbf{Results:}
The results construct a catalog of \finalpool \textit{Extract Method} approaches indicating that several techniques have been proposed in the literature. Our results show that: 
\textcolor{black}{(i) 38.6\% of \textit{Extract Method} refactoring studies
primarily focus on addressing code clones;
(ii) Several of the \textit{Extract Method}  tools incorporate the developer’s involvement in
the decision-making process when applying the method extraction, and (iii)  the existing benchmarks are heterogeneous and do not contain
the same type of information, making standardizing them
for the purpose of benchmarking difficult.}


\textbf{Conclusions:} Our study serves as an \say{index} to the body of knowledge in this area for researchers and practitioners in determining the
\textit{Extract Method} refactoring approach 
 that is most appropriate for their needs. Our findings also empower the
community with information to guide the future development of refactoring tools.

\end{abstract}

\begin{IEEEkeywords}
extract method, refactoring, quality, systematic literature review
\end{IEEEkeywords}
}
\maketitle
\IEEEdisplaynontitleabstractindextext
\IEEEpeerreviewmaketitle

\begin{figure*}[htbp]
 	\centering
 	\includegraphics[width=\textwidth]{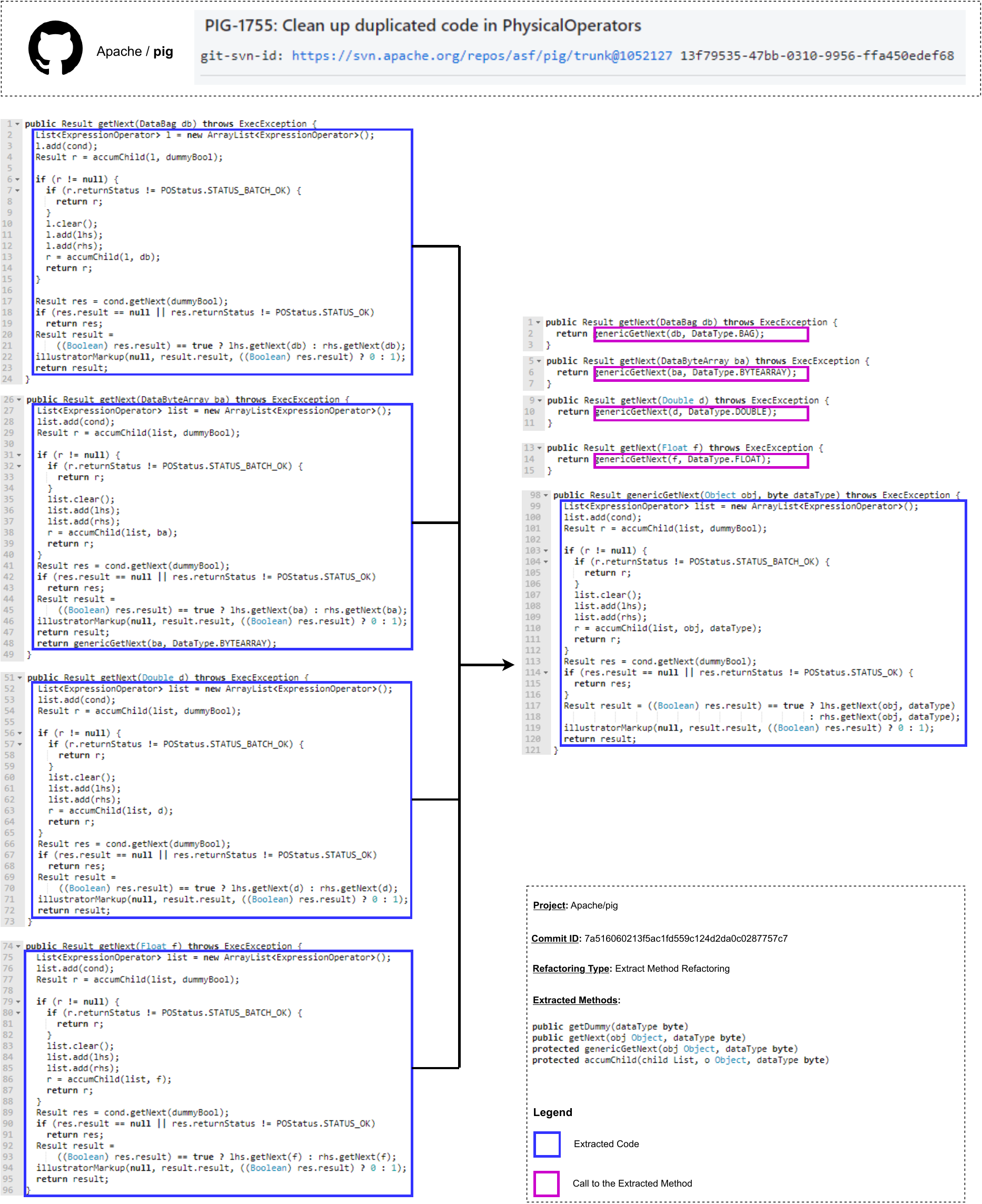}
 	\caption{\textcolor{black}{Sample example of \textit{Extract Method} refactoring \cite{pig}.}}
 	\label{fig:example}
\end{figure*}

\section{Introduction}
\label{Section:Introduction}

\IEEEPARstart{R}{efactoring} is the art of restructuring code to improve it without changing its external behavior \cite{fowler2018refactoring}. One of the basic building blocks of refactoring is \textit{Extract Method}, \ie the process of moving a fragment of code from an existing method into a new method with a name that explains its behavior. Method extraction is one of the main refactorings that were defined when this area was established \cite{griswold1993automated}, as it is a common response to the need of keeping methods concise and modular, and reducing the spread of shared responsibilities. Furthermore, \textit{Extract Method} serves as a bridge to facilitate more complex refactorings \cite{zarras2015navigating}. \textcolor{black}{\textit{Extract Method} is widely employed by developers across various  systems\footnote{Based on \texttt{JDeodorant} tool usage statistics: "https://users.encs.concordia.ca/~nikolaos/"}. It represents approximately 49.6\% of the total refactorings recommended, as shown by JDeodorant \cite{Jdeodorant}, one of popular tools that support \textit{Extract Method} refactoring. Moreover, open-source developers \cite{Silva2016why,tsantalis2013multidimensional,murphy2008breaking,charalampidou2018structural,charalampidou2016identifying,thy15adventure,murphy2011we,alomar2019can} and industry professionals \cite{van2021data} consider it a critical refactoring operation. The popularity of this refactoring is inherited from its multifaceted utility that can be used for a myriad of reasons, such as removal of duplicate code \cite{yoshida2019proactive,alcocer2020improving,hotta2010duplicate,higo2008metric}, extraction of reusable methods \cite{alomar2022refactoring,Silva2016why,alomar2020developers}, wrapping older method signatures \cite{Silva2016why}, decomposition of long or complex structures \cite{yang2009identifying,morales2017use,tiwari2022identifying,khomh2012exploratory,palomba2014they,palomba2018diffuseness,oizumi2020recommending}, and support of code testability \cite{cinneide2011automated,harman2011refactoring}. This wide variety of usage scenarios shows why method extraction is considered the\textit{ Swiss Army knife} of refactoring operations \cite{hora2020characteristics}. One of the typical rationales behind method extraction is the removal of duplicate code instances, which we can extract from a real-world case. In this case, the committer has documented the cleaning up of duplicate code. A closer inspection of the code changes, illustrated in Figure \ref{fig:example},  reveals the elimination of code duplication in four methods (\ie \texttt{getDummy(dataType byte)}, \texttt{getNext(obj Object, dataType byte}, \texttt{genericGetNext(obj Object, dataType byte), and accumChild(child List, o Object, dataType byte)}, where four duplicates are extracted into one separate method (\ie \texttt{genericGetNext(Object obj, byte dataType)} and then replaced with calls to the newly extracted method.}

\textcolor{black}{Given its popularity and the diversity of its usage scenarios, modern Integrated Development Environments (IDEs), such as IntelliJ IDEA, PyCharm, Eclipse, and Visual Studio offer the \textit{Extract Method} refactoring as a built-in feature, to support the correctness of code transformation and its behavior preservation. However, the built-in feature only supports the \textit{automation} of the refactoring and not the \textit{recommendation} of opportunities to apply it.} Therefore, various research projects focused on recommending method extraction, by identifying refactoring opportunities, such as making code more reusable \cite{alomar2022refactoring,Silva2016why,alomar2020developers}, removing duplicate code \cite{yoshida2019proactive,alcocer2020improving,hotta2010duplicate,higo2008metric}, improving testability through smaller test methods \cite{cinneide2011automated,harman2011refactoring}, and segregating multiple functionalities \cite{yang2009identifying,morales2017use,tiwari2022identifying,khomh2012exploratory,palomba2014they,palomba2018diffuseness,oizumi2020recommending}. Some of these studies have also implemented their solutions in tools and plugins. 

\textcolor{black}{Despite the existence of built-in IDE features, and tools, several surveys report a general reluctance of developers to adopt them \cite{kim2014empirical,Silva2016why,golubev2021one,ivers2022industry,alves2017refactoring}. In fact, surveys show that developers tend to manually extract methods despite the associated effort and error-proneness 
 \cite{golubev2021one}.} \textcolor{black}{Existing research assumes that practitioners have a clear and common understanding of the intent behind method extraction, since it focuses on improving the accuracy of identifying refactoring opportunities. Yet, a recent investigation of Stack Overflow posts, related to \textit{Extract Method}, outlines how developers are asking how to perform refactoring, whether there is tool support, and how to avoid any side effects \cite{hora2020characteristics}.} 
 Bridging the gap between the state-of-the-art and the state-of-the-practice starts with understanding the \textit{intent} that drives primary studies (PSs) to identify refactoring opportunities and the extent to which they support its execution. In fact, cataloging these studies can facilitate their adoption by developers. Therefore, this paper systematically maps existing research in the recommendation of \textit{Extract Method} refactoring from six main dimensions: 
 

 \begin{itemize}
  \item \textcolor{black}{\textbf{Intent:} refers to the motivation behind the need for a method to be extracted, \eg duplicate code removal.}
  \item \textcolor{black}{\textbf{Code Analysis:} refers to the type of source code analysis, \eg lexical and semantic code analysis.}
  \item \textcolor{black}{\textbf{Code Representation:} refers to the underlying code representation being used during the extraction, \eg source code and AST.}
  \item \textcolor{black}{\textbf{Detection:} refers to the automation degree to which a refactoring opportunity is detected, \eg manual and fully-automated.} 
  \item \textcolor{black}{\textbf{Execution:} refers to the automation degree to which a refactoring opportunity is executed, \eg manual and fully-automated} 
  \item \textcolor{black}{\textbf{Validation Method:} refers to the approaches that have been suggested for evaluation method extraction, \eg case study, and experiment.}
\end{itemize}
 

 Another interesting investigation relates to the existing toolset implemented by researchers. We further classify them based on various characteristics, including their target language, availability, types of validation, etc.

Since little is known about the existing literature on \textit{Extract Method} refactoring, this SLR serves as a comprehensive review of the body of \textcolor{black}{knowledge on} this topic to analyze existing techniques, 
 and their associated programming languages. The analysis of such a wide variety of methods leads to the development of categorization and reveals areas of potential improvements. Therefore, when defining our research questions, we follow established guidelines in systematic literature review studies \cite{Kitchenham07guidelinesfor,wohlin2014guidelines,petersen2008systematic}. The motivation behind each question is as follows.
\begin{itemize}
    \item \textbf{RQ$_1$: \RQone} 
           \textcolor{black}{We pose this RQ to study current approaches for \textit{Extract Method}, and to get an overview of the existing approaches and their characteristics. Accordingly, for each surveyed study, we collect information about six main dimensions, together with any associated tools.} 
    \item \textbf{RQ$_2$: \RQtwo}
           This RQ dives deeper into the characteristics of the tools. It outlines how they were implemented, maintained, and validated. 
           \item \textcolor{black}{\textbf{RQ$_3$: \RQthree} This RQ investigates the datasets, and benchmarks, which refers to systems and system artifacts, that are chosen and used for evaluating and validating the extraction of methods, and its results.}

\end{itemize}

The main contributions of this paper are summarized as follows:
\begin{itemize}
 \item We conduct the first SLR to review \textit{Extract Method} refactoring, and classifying its corresponding studies from various dimensions.
 
 \item We explore the existing toolset and \textcolor{black}{benchmarks} generated by these studies. \textcolor{black}{We provide a one-stop-shop website that links to all the tools and datasets that we were able to recover from the studies\footnote{\url{https://refactorings.github.io/em-slr/}}}. 
 
 \item We provide practical implications of our findings for researchers, developers, tool builders, and educators.
 
\end{itemize}

The remainder of this paper is organized as follows: Section \ref{Section:Background} reviews existing studies related to systematic reviews of refactoring. Section \ref{Section:methodology} outlines our empirical setup in terms of search strategy, study selection, and data extraction. Section \ref{Section:Result} discusses our findings, while the \textcolor{black}{research implications are} discussed in Section \ref{Section:Reflection}. Section \ref{Section:Threats} captures threats to the validity of our work before concluding with Section \ref{Section:Conclusion}.



\section{Related Work}
\label{Section:Background}

\begin{table}[ht!]
\begin{center}
\caption{Refactoring-related SLRs in related work.}
\label{Table:RelatedWorkSLR}
\begin{adjustbox}{width=0.5\textwidth,center}
\rowcolors{2}{white}{gray!25}
\begin{tabular}{lllc}\hline
\toprule
\bfseries Study & \bfseries Year  & \bfseries Focus & \bfseries No of PSs \\
\midrule
Zhang \etal \cite{Zhang:2011:CBS:1967084.1967086} & 2011 & Bad smells \& refactoring & 39 \\

Abebe \& Yoo \cite{article2} & 2014 & Refactoring trends \& challenges & 58 \\


AlDallal \cite{ALDALLAL2015231} & 2015 & Refactoring identification & 47 \\

Singh \& Kaur \cite{SINGH2017} & 2017 & Refactoring identification & 238 \\

AlDallal \& Abdin \cite{7833023} & 2017 & Impact of refactoring on quality & 76 \\

Mariani \& Vergilio \cite{MARIANI201714} & 2017 & Search-based refactoring & 71 \\

Baqais \& Alshayeb \cite{baqais2020automatic} & 2020 & Automatic refactoring & 41 \\

Lacerda \etal \cite{lacerda2020code} & 2020 & Code smells \& refactoring & 40 \\

Abid \etal \cite{abid202030} & 2020 & Refactoring research efforts & 3183 \\

AlOmar \etal \cite{alomar2021preserving} & 2021 & Refactoring behavior preservation & 28 \\
\textbf{This work} & & \textbf{\textit{Extract Method} refactoring} & \textbf{83} \\
\bottomrule
\end{tabular}
\end{adjustbox}
\end{center}
\end{table}

Zhang \etal \cite{Zhang:2011:CBS:1967084.1967086} conducted a systematic literature review (SLR) on 39 studies on bad code smells. They discussed these studies based on various aspects including the goals of the studies, the type of code smells, the approaches to detect code smells, and finally, their refactoring opportunities. \textcolor{black}{Their main finding shows that \textit{Duplicated Code} and \textit{Long Method} are among the most studied code smells. Furthermore, they found that nearly 49\% of the primary studies aim to improve tools to detect code smells, while only 15\% focus on enhancing the current knowledge of refactoring code smells.}
Later, Abebe and Yoo \cite{article2} conducted another systematic review of 58 studies to reveal software refactoring trends, opportunities, and challenges. Their classification helped guide researchers to address the crucial issues in software refactoring. \textcolor{black}{The authors pointed out that one of the gaps in refactoring research is the lack of a refactoring tool that provides custom refactoring for all specific user needs.} After that, 
  AlDallal \cite{ALDALLAL2015231} conducted an SLR of 47 PSs published on identifying refactoring opportunities in object-oriented code. AlDallal's review classified PSs based on the considered refactoring scenarios, the approaches to determine refactoring candidates, and the datasets used in the existing empirical studies. \textcolor{black}{In their study, \textit{Extract Method} refactoring is used in refactoring identification approaches, \ie quality metrics-oriented, precondition-oriented, clustering-oriented, graph-oriented, and code-slicing-oriented approaches}. In the following SLR work by AlDallal and Abdin \cite{7833023}, they discussed 76 PSs and classified them based on refactoring quality attributes of object-oriented code. \textcolor{black}{Their finding shows that the authors of the PSs studied the
impact of the \textit{Extract Method}
 refactoring on quality much more frequently, and was considered by 11.8\% or
more of the PSs.} Thereafter, Singh and Kaur \cite{SINGH2017} performed an SLR as an extension of AlDallal's SLR \cite{ALDALLAL2015231} where they analyzed 238 research items in code smell detection and its refactoring opportunities to address some research questions left open in AlDallal's SLR. \textcolor{black}{Their finding reveals that \textit{Extract Method} refactoring was used in metric-based detection techniques.} Baqais and Alshayeb \cite{baqais2020automatic} conducted a systematic literature review on automated software refactoring. In their review, they analyzed 41 studies that propose or develop different automatic refactoring approaches,  \textcolor{black}{finding that \textit{Extract Method} used in precondition-based approaches.}

Other studies focus on search-based refactoring where search techniques are used to identify refactoring recommendations. Mariani and Vergilio \cite{MARIANI201714} systematically reviewed 71 studies and classified them based on the main elements of search-based refactoring, including artifacts used, encoding and algorithms used, search technique, metrics addressed, available tools, and conducted evaluation. Mariani and Vergilio classified the selected PSs into five general categories related to behavior preservation methods. These categories involved (1) Opdyke's function \cite{Opdyke:1992:ROF:169783}, (2) Cinn{\'e}ide's function \cite{cinneide2001automated}, (3) domain-specific, (4) no evidence of behavior preservation, and (5) do not mention the method. \textcolor{black}{One of their main takeaways is the need for search-based approaches to explore the need to achieve fully automated approaches for refactoring.} 
Lacerda \etal \cite{lacerda2020code} performed a tertiary systematic literature review of 40 secondary studies to identify the main observations and challenges on code smell and refactoring. Their finding shows that code smells and refactoring strongly correlate with quality attributes. \textcolor{black}{They concluded that few refactoring tools exist, and some are obsolete. There is an opportunity to propose
and improve \textit{Extract Method} refactoring tools, especially tools to predict and evaluate the effects of refactoring.} 
Abid \etal \cite{abid202030} analyzed the results of 3,183 primary studies on refactoring covering the last three decades to offer a comprehensive literature review of existing refactoring research studies.  The authors derived a taxonomy focused on five key aspects of refactoring including refactoring lifecycle, artifacts affected by refactoring, refactoring objectives, refactoring techniques, and refactoring evaluation.  \textcolor{black}{They highlight the need to validate refactoring techniques and tools using industrial systems to bridge the gap between academic research and industry's research needs.}

AlOmar \etal \cite{alomar2021preserving} conducted a systematic literature mapping to identify behavior preservation approaches in software refactoring. Their key finding reveals the variety of formalisms and techniques such as developing automatic refactoring safety tools and performing a manual source code analysis. However, researchers are biased toward using precondition-based and testing-based approaches although there are other techniques (\eg graph-based) that have some potential
and perhaps \textcolor{black}{they are effective} for specific problems that have not yet been well explored. \textcolor{black}{Further, the authors found that \textit{Extract Method} refactoring is one of the most widely used refactoring operations in PSs to demonstrate behavior preservation.}

Table~\ref{Table:RelatedWorkSLR} summarizes existing SLRs on software refactoring. Overall, we observe that all the above-mentioned studies focus on either (1) detecting refactoring opportunities through the optimization of structural metrics or the identification of design and code defects,  (2) automating the generation and recommendation of the most optimal set of refactorings to improve the system's design while minimizing the refactoring effort, so that developers still can recognize their own design, or (3) demonstrating comprehensive literature review of existing refactoring research studies and the concept of behavior preservation. Our work differs from these studies, as our SLR focuses primarily on collecting and summarizing specifically \textit{Extract Method} refactoring techniques, the \say{Swiss army knife of refactorings} \cite{Silva2016why,tsantalis2013multidimensional} with an in-depth analysis. To the best of our knowledge, no previous work has conducted a comprehensive SLR pertaining to \textit{Extract Method} techniques in software refactoring.

\section{Study Design}
\label{Section:methodology}
\begin{figure*}[t]
 	\centering
 	\includegraphics[width=1.0\textwidth]{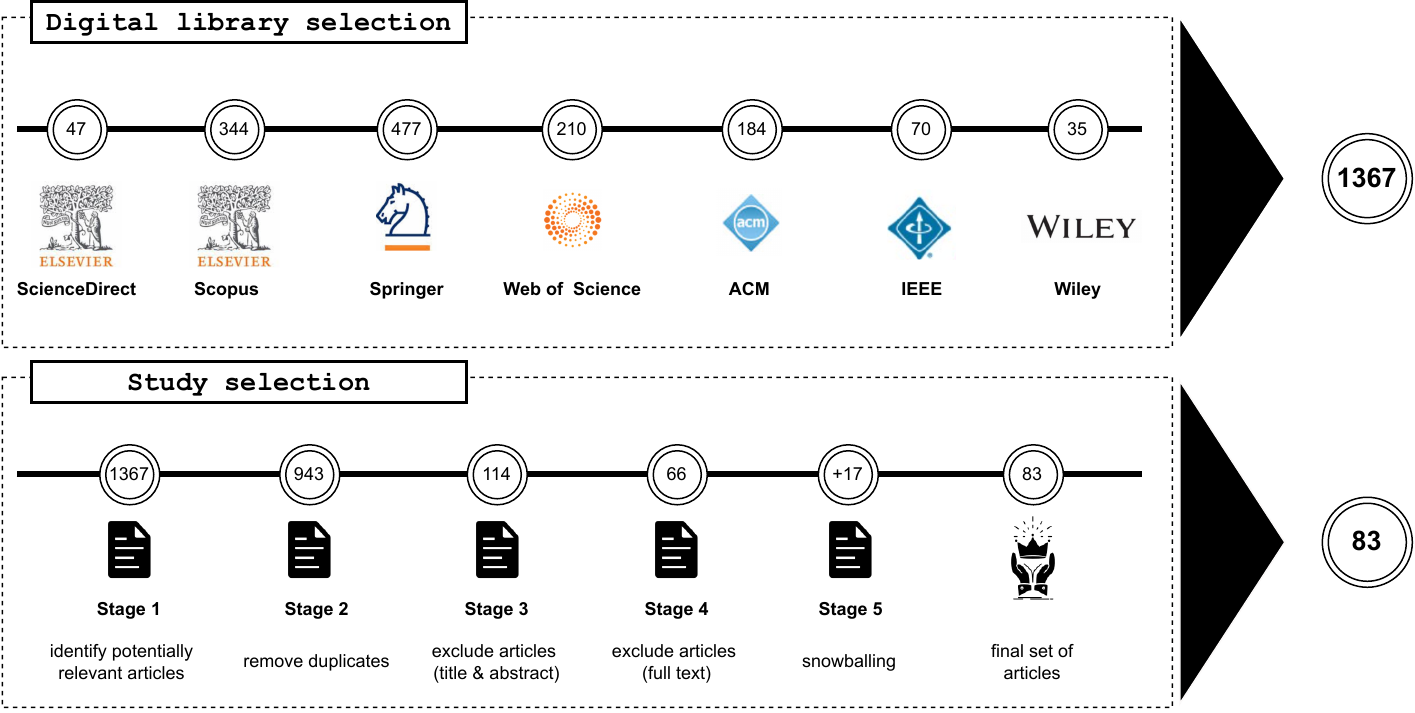}
 	\caption{\textcolor{black}{Literature search process.}}
 	\label{fig:approach}
\end{figure*}

This SLR aims to explore the landscape of approaches and tools that recommend the \textit{Extract Method} refactoring. Based on established guidelines \cite{Kitchenham07guidelinesfor,wohlin2014guidelines,kitchenham2004procedures,brereton2007lessons,kitchenham2009systematic}, we performed the SLR in three main phases: planning, reviewing, and reporting the review. Creating a protocol is a major step when conducting an SLR \cite{Kitchenham07guidelinesfor}. \textcolor{black}{The planning phase involves identifying the need for a review and the development of a review protocol (described in Section \ref{sec:planning}). The review phase encompasses the selection of primary studies, the assessment of the study, data extraction, and data synthesis (described in Sections \ref{sec:selection} and \ref{sec:analysis}). Finally, the reporting phase emphasizes recording the review, which involves observing documents, and presenting the obtained results (described in Section \ref{Section:Result}).} 

\subsection{Survey Planning}
\label{sec:planning}
\textcolor{black}{The planning phase highlights the research motivation that leads to the development of research questions.}

\subsubsection{Identifying the need for a Systematic Literature Review} \textcolor{black}{The absence of comprehensive and current secondary research that delves into the \textit{Extract Method} underscores the need for a comprehensive Systematic Literature Review (SLR). While there have been SLRs in the field of refactoring, their focus remains confined to the automation of refactoring, the impact of refactoring on quality, detection of code smells and trends, challenges, and application of refactoring, which none specializes in \textit{Extract Method}. Thus, the core motivation behind carrying out this SLR is to}:
\begin{itemize}
\item Collect the body of knowledge of \textit{Extract Method} refactoring approaches in the research literature.
\item Combine and analyze the reported findings regarding \textit{Extract Method} approaches. 
\item Identify open issues in existing research.
\end{itemize} 

\subsubsection{Specifying the research questions} \textcolor{black}{During the process of conducting an SLR,  it is of paramount importance to pinpoint pertinent research questions that have the potential to provide clear answers. We identified  three such research questions:}

\begin{itemize}
    \item \textcolor{black}{\textbf{RQ$_1$: \RQone}} 
    \item \textcolor{black}{\textbf{RQ$_2$: \RQtwo}}
    \item \textcolor{black}{\textbf{RQ$_3$: \RQthree}}        

\end{itemize}

\subsection{Primary Studies Selection}
\label{sec:selection}
\textcolor{black}{In alignment with the research questions, we extracted the initial terms that encapsulated the research topic. Referring to previous reviews of the literature within the field, we developed search keywords incorporating synonyms and related terms.}

\subsubsection{Search strategy}
Similar to Fernandes \etal \cite{fernandes2016review}, we performed an automatic search in seven electronic
data sources  to find relevant studies, including ScienceDirect\footnote{https://www.sciencedirect.com/}, Scopus\footnote{https://www.scopus.com}, Springer Link\footnote{https://link.springer.com/}, Web of Science\footnote{https://webofknowledge.com/},  ACM Digital Library\footnote{https://dl.acm.org/}, IEEE Xplore\footnote{https://ieeexplore.ieee.org/}, and Wiley\footnote{https://onlinelibrary.wiley.com/}. TextBox \ref{txt:search_string} shows our search string in these search engines.  
\begin{TextBox}[h]
\centering
\fbox{\begin{minipage}{26em}
\textcolor{black}{((extract method OR  extract-method OR method extract* OR method-extract* OR extract function OR extract-function OR function extract* OR function-extract* OR split method OR split-method OR method split* OR method-split* OR split function OR split-function OR function split* OR function-split* OR  separat* method OR separat*-method OR  method separat* OR method-separat* OR separat* function OR separate-function OR function separat* OR function-separat*) AND (long method OR long function OR large method OR large function OR duplicat* code OR code duplicat* OR code clone OR code bad smell OR code smell OR bad smell OR antipattern OR anti-pattern OR design defect OR design flaw) AND (refactor*) AND (approach OR tool OR technique))}
\end{minipage}}
\captionof{TextBox}{Search string.\label{txt:search_string}} 
\end{TextBox}

The strategy to construct our search keywords is as follows:
\begin{itemize}
\item Derive the main terms from research questions and terms considered in the relevant papers.
\item Include alternative spellings for major terms.
\item Combine possible synonyms and spellings of the main terms using Boolean OR operators and then combine the main terms using the Boolean AND operators. 
\end{itemize}

\begin{figure}[t]
 	\centering
 	\includegraphics[width=1.0\columnwidth]{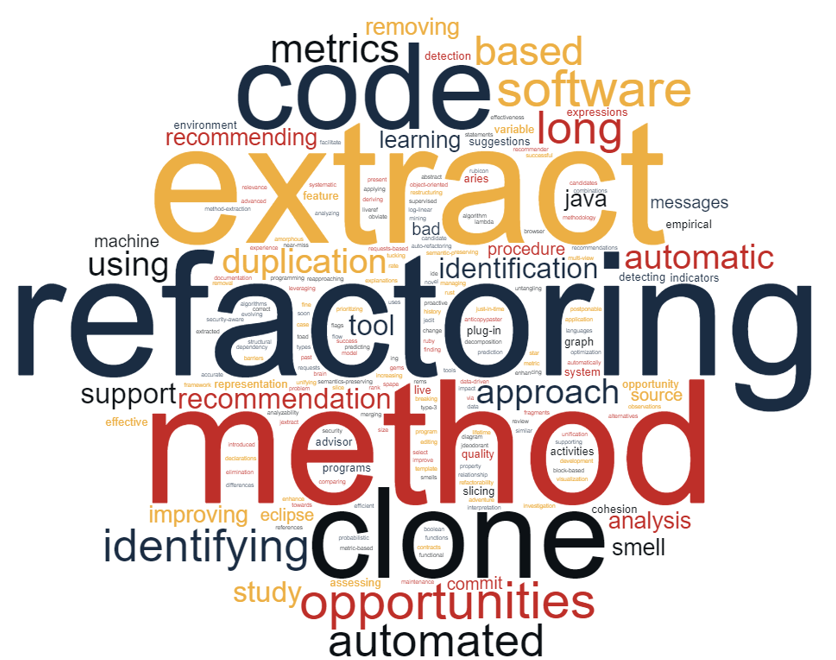}
 	\caption{\textcolor{black}{Word cloud of paper titles of primary studies.}}
 	\label{fig:wordcloud}
\end{figure}

These search keywords are applied to titles, abstracts, and keywords. To verify the validity of the search string, we manually double-checked a few articles from each of the \textcolor{black}{seven} digital libraries, similar to Garousi and Mäntylä \cite{GAROUSI2016195}. 
 \textcolor{black}{Also, during the review of this manuscript, reviewers pointed out a set of keywords whose incorporation helped with revealing more studies that were finally included.} 
\textcolor{black}{To get a high-level picture of the covered topics, we generated a word cloud of paper titles, as depicted in Figure \ref{fig:wordcloud}.}

\subsubsection{Study selection}
To collect the PSs, we adapted the search process of AlDallal and Abdin \cite{7833023} \cite{7833023} 
and conducted a five-phased process. \textcolor{black}{Literature publications were eliminated based on the defined inclusion and exclusion criteria to filter our irrelevant articles.}

\textit{Inclusion criteria (IC):}

\textcolor{black}{The selected studies must satisfy all the following inclusion criteria:}
\begin{itemize}
\item \textcolor{black}{The article must be published in peer-reviewed venues before \textcolor{black}{August 26, 2023}.}
\item \textcolor{black}{The article must report an approach to recommend \textit{Extract Method} refactoring.}
\end{itemize}

\textit{Exclusion criteria (EC):}

\textcolor{black}{Papers are excluded if satisfying any of the exclusion criteria, as follows:}
\begin{itemize}
\item \textcolor{black}{The study is a position 
paper, abstract, blog,  editorial, keynote, tutorial, book, patent, or panel discussion.}
\item \textcolor{black}{The study is not written in English.}
\end{itemize}

\textcolor{black}{Regarding the second inclusion criteria, we only considered PSs that reported an approach to recommend \textit{Extract Method} refactoring. We excluded any other articles that provided a broad explanation of the concept of \textit{Extract Method} refactoring.} 

\textcolor{black}{\textbf{Stage 1: Identification of potentially relevant articles.}} In this first stage of the selection process, shown in Figure~\ref{fig:approach}, we searched seven digital libraries 
 for potentially related articles. Our criteria included applying our predefined search string to the title, abstract, and keyword fields. The results of this search were not limited to specific venues. Searching through the seven digital libraries resulted in a total of \initialpool publications in the literature. We performed the initial screening of the articles to reduce the possibility of including irrelevant articles.

\textcolor{black}{\textbf{Stage 2: Removal of duplicates.}}  \textcolor{black}{By merging the results obtained from the search platforms, we remove duplicate publications, books, and reports, which resulted in a total of \noduplicate literature publications.}

\textcolor{black}{\textbf{Stage 3: Exclusion of articles based on title and abstract.}} 
It is important to consider the abstracts at this stage because the titles of some articles could be misleading. Inclusion and exclusion rules were applied at this stage to all retrieved studies. This elimination process reduced our set of results to \titleabstract publications in the literature. \textcolor{black}{When a determination cannot be reached solely based on the title and abstracts, the studies are promoted to the next stage.}

\textcolor{black}{\textbf{Stage 4: Exclusion of articles based on full text.}} To obtain the relevant PSs, the identified papers in Stage 3 were reviewed. Literature reviews were eliminated based on defined exclusion and inclusion rules. This process resulted in a total of \fulltext literature publications that were included in this study. 

\textcolor{black}{\textbf{Stage 5: Snowballing.}} To maximize the search coverage of all relevant papers, we also performed the snowballing technique \cite{wohlin2014guidelines} on \fulltext papers already in the pool. \textcolor{black}{Using snowballing, we extracted \reference references from the reference section of the studies, and extracted studies citing the \fulltext selected studies. We combined the results and filtered out duplicate records, along with books, and non-peer reviewed studies. Then, we compare this set with \noduplicate primary studies obtained from Stage 2 to further refine the studies.} This step resulted in the addition of \snowballing additional papers, \textcolor{black}{where some of them did not explicitly mention the recommendation of \textit{Extract Method} in their titles and abstracts.} The updated pool size increased to \finalpool.

\subsection{Study Quality Assessment}
\textcolor{black}{To assess the quality of PSs, we followed the guidelines proposed in  \cite{Kitchenham07guidelinesfor,li2021understanding,dybaa2008empirical}. We chose three quality assessment questions that could be applicable to all PSs, and each PS is evaluated against three questions within three dimensions of study quality (\ie objective, method, and coverage of the studies). The corresponding questions are as follows: Q1) \textit{Does the study’s primary objective explicitly focus on the \textit{Extract Method} refactoring?}; Q2) \textit{Does the study include structured and preferably automatic
or semi-automatic \textit{Extract Method} approaches?}; and Q3) \textit{Does the
study sufficiently describe the \textit{Extract Method} technique,
algorithm, and evaluation?}. These questions are implicitly used
in the above refinement stages. If a PS passes these quality criteria, we believe that a PS has valuable information for SLR. The answer to each of these questions is either \say{Yes}, \say{Partially}, or \say{No} with numerical values of 1, 0.5, or 0, respectively. If the questions did not apply to the context of a PS, they were not evaluated. The overall quality of each PS is calculated by summing up the scores of the applicable questions. In general, all the published articles in the accepted literature scored well on the quality assessment questions.}

\subsection{Data Extraction, Categorization, and Analysis}
\label{sec:analysis}

To determine the attribute(s) of the classification dimension \cite{KITCHENHAM20132049,lenarduzzi2021systematic}, we screened the full texts of the PSs and identified the attribute(s) of that dimension. We used attribute(s) generalization and refinement to derive the final map, similar to \cite{GAROUSI2016195}. Specifically, we analyzed the PSs to create a comprehensive high-level list of themes, extracted from a thematic analysis, based on guidelines provided by Cruzes~\etal~\cite{cruzes2011recommended}. Thematic analysis is among the most used methods in Software Engineering literature~\cite{Silva2016why,alomar2022code,alomar2021refactoring}, for identifying and recording patterns (or \say{themes}) within a collection of descriptive labels, which we call \say{codes}. For each PS, we proceeded with the analysis using the following steps: \textit{i}) Initial reading of the PSs; \textit{ii}) Generating initial codes (\ie labels) for each PS; \textit{iii}) Translating codes into themes, sub-themes, and higher-order themes; \textit{iv}) Reviewing the themes to find opportunities for merging; \textit{v}) Defining and naming the final themes, and creating a model of higher-order themes and their underlying evidence. 

\textcolor{black}{
Inspired by previous studies \cite{roy2009comparison,perez2011classification}, we initiated our study by adopting existing taxonomies to categorize PSs. To carry out the manual coding of PSs, we used a spreadsheet application equipped with tagging capabilities. This spreadsheet provided the annotators with the following information: (1) the paper title and study link, (2) why \textit{Extract Method} is performed (\ie intent), (3) the type of source code analysis (\ie code analysis), (4) the underlying code representation used during the extraction (\ie representation), (5)  the automation degree of detecting the refactoring opportunity, (6) the automation degree of executing the recommended refactoring, and (7) the type of experiments carried out to validate the method. When creating our customized classification dimensions, annotators could select from preexisting tags in a drop-down menu or create a new one if none of the existing tags fits the specific case (\ie  each annotator had the flexibility to assign one or more tagging items).}


The above-mentioned steps were performed independently by two authors. One author performed the labeling of PSs independently of the other author, who was responsible for reviewing the currently drafted themes. At the end of each iteration, the authors met and refined the themes to reach a consensus. It is important to note that the approach is not a single-step process. As the codes were analyzed, some of the first cycle codes were subsumed by other codes, relabeled, or dropped altogether. \textcolor{black}{As the two authors progressed in translating the themes, there was some reorganization, refinement, and reclassification of the data into different or new codes.} \textcolor{black}{For example, we aggregated, into \say{Intent}, the preliminary categories \say{duplicated code}, \say{code clone}, \say{long method}, and \say{separation of concerns}. We used the thematic analysis technique to address RQ$_1$ and RQ$_2$.} 

\subsection{\textcolor{black}{Final Primary Studies Selection}}

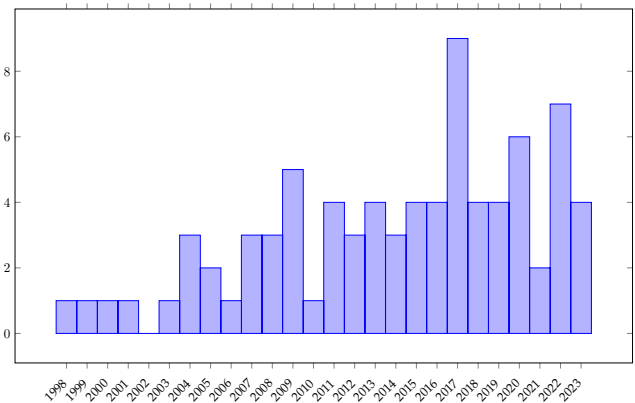
\begin{figure}[h]  
\centering 
\begin{tikzpicture}
\begin{scope}[scale=0.5]
\begin{axis}[
    ybar,
    height=11cm,
    width=18cm,
    bar width=0.55cm, 
    xticklabels from table={\dataA}{Category}, 
    xtick=data,
    x tick label style={
      rotate=45,
      anchor=east, 
      xshift=-1.5mm, yshift=-2mm
    },
    legend style={
      at={(0.5,-0.2)},
      anchor=south,
      legend columns=-1
      },
]

\addplot coordinates {
    (1998,1) 
    (1999,1) 
    (2000,1)
    (2001,1) 
    (2002,0) 
    (2003,1) 
    (2004,3) 
    (2005,2)
    (2006,1)
    (2007,3)
    (2008,3)
    (2009,5)
    (2010,1)
    (2011,4)
    (2012,3)
    (2013,4)
    (2014,3)
    (2015,4)
    (2016,4)
    (2017,9)
    (2018,4)
    (2019,4)
    (2020,6)
    (2021,2)
    (2022,7)
    (2023,4)
};
\end{axis}
\end{scope}
\end{tikzpicture}
\caption{\textcolor{black}{Distribution of primary studies by year.}} 
\label{fig:Distribution_PSs_by_year}
\end{figure}

The research method discussed in Section~\ref{Section:methodology} resulted in \finalpool relevant PSs. The main venues for these relevant PSs are presented in Table~\ref{Table:Publication_Sources}. The PSs were published in 55 different sources, including journals, conferences, and workshops. The list specifically includes 12 journals, 37 conferences, and 8 workshops. The first relevant article was published in a journal in 1998, whereas the most recent one was published in 2023. The number of literary papers published in journals, conferences, and workshops combined, is presented in Figure~\ref{fig:Distribution_PSs_by_year}. This figure illustrates a trend that began in 2017, resulting in a higher number of studies conducted between 2017 and 2023 compared to the total of studies published before 2017. This growing interest in this refactoring incites further research to improve its adoption in practice. 

\begin{table}[h]
\begin{center}
\caption{\textcolor{black}{Publication venues.}}
\label{Table:Publication_Sources}
\begin{adjustbox}{width=0.5\textwidth,center}
\begin{tabular}{llll}\hline
\toprule
\bfseries Publication Venue & \bfseries PSs  \\
\midrule
 \cellcolor{gray!25}Symposium on Software Reusability & \cellcolor{gray!25}\cite{maruyama2001automated}\\
 International Conference on Software Engineering  & \cite{murphy2008breaking,antezana2019toad,maruyama2017tool,mazinanian2016jdeodorant,tsantalis2017clone,meng2015does} \\ 
 \cellcolor{gray!25}Conference on Software Maintenance and Reengineering & \cellcolor{gray!25}\cite{tsantalis2009identification}  \\
 Journal of Systems and Software & \cite{tsantalis2011identification,bian2013spape,shahidi2022automated} \\
 \cellcolor{gray!25}Asia-Pacific Software Engineering Conference & \cellcolor{gray!25}\cite{yang2009identifying,choi2018investigation}  \\
Workshop on Refactoring Tools & \cite{kanemitsu2011visualization,sharma2012identifying,abadi2009fine,juillerat2007improving,abadi2008re}   \\
\cellcolor{gray!25}International Conference on Program Comprehension & \cellcolor{gray!25}\cite{silva2014recommending,cui2023rems,komondoor2003effective}    \\
 Agile Processes in Software Engineering and Extreme Programming & \cite{arcelli2015duplicated} \\
\cellcolor{gray!25}Transactions on Software Engineering & \cellcolor{gray!25}\cite{charalampidou2016identifying,aniche2020effectiveness,abid2020does,tsantalis2015assessing}   \\
 International Conference on Software Quality &\cite{haas2016deriving,haas2017learning} \\
\cellcolor{gray!25}International Symposium on Software Reliability Engineering & \cellcolor{gray!25}\cite{xu2017gems}  \\
 International Conference on Software Maintenance and Evolution & \cite{yue2018automatic,palit14automatic}  \\
 \cellcolor{gray!25}International Workshop on Refactoring & \cellcolor{gray!25}\cite{yoshida2019proactive}  \\
IEEE Access &  \cite{sheneamer2020automatic} \\
\cellcolor{gray!25}Symposium on the Foundations of Software Engineering & \cellcolor{gray!25}\cite{van2021data}  \\
 Innovations in Software Engineering Conference & \cite{tiwari2022identifying}  \\ 
 \cellcolor{gray!25}International Conference on Automated Software Engineering  & \cellcolor{gray!25}\cite{alomar2022anticopypaster,fernandes2022liveref} \\
Information and Software Technology &\cite{lakhotia1998restructuring,tairas2012increasing,alomar2023just}  \\
\cellcolor{gray!25}Science of Computer Programming & \cellcolor{gray!25}\cite{alcocer2020improving} \\
Conference on Software: Theory and Practice & \cite{silva2015jextract} \\
\cellcolor{gray!25}International Conference on the Art, Science, and Engineering of
Programming & \cellcolor{gray!25}\cite{fernandes2022live} \\
Computer Software and Applications Conference & \cite{imazato2017finding} \\
\cellcolor{gray!25}International Journal of Software Engineering and Knowledge Engineering & \cellcolor{gray!25}\cite{kaya2017identification} \\
International Conference on Software Engineering and Knowledge Engineering & \cite{kaya2013identifying} \\
\cellcolor{gray!25}Automated Software Engineering Journal & \cellcolor{gray!25}\cite{alomar2022documentation} \\
Machine Learning with Applications & \cite{nyamawe2022mining} \\
\cellcolor{gray!25}Empirical Software Engineering & \cellcolor{gray!25}\cite{nyamawe2020feature} \\
International Requirements Engineering Conference & \cite{nyamawe2019automated} \\
\cellcolor{gray!25}Algorithms & \cellcolor{gray!25}\cite{sagar2021comparing} \\
International Conference on Software Analysis, Evolution and Reengineering & \cite{krasniqi2020enhancing,ettinger2017efficient,ettinger2016duplication} \\
\cellcolor{gray!25}International Federation for Information Processing & \cellcolor{gray!25}\cite{vittek2007c++} \\
Conference on Object-oriented programming systems and applications & \cite{corbat2007ruby,cousot2012abstract,thy15adventure} \\
\cellcolor{gray!25}IEICE Transactions on Information and Systems & \cellcolor{gray!25}\cite{meananeatra2018refactoring} \\
International Conference on Computer and Communications & \cite{xu2017log} \\
\cellcolor{gray!25}IASTED Conf. on Software Engineering and Applications & \cellcolor{gray!25}\cite{higo2004aries} \\
ACM SIGSOFT Software Engineering Notes & \cite{higo2005aries} \\
\cellcolor{gray!25}OOPSLA workshop on Eclipse technology eXchange & \cellcolor{gray!25}\cite{o2005star} \\
International Conference on Product Focused Software Process Improvement & \cite{higo2004refactoring} \\
\cellcolor{gray!25}Journal of Software Maintenance and Evolution: Research and Practice & \cellcolor{gray!25}\cite{higo2008metric} \\
International Conference on Soft Computing Techniques and Engineering Application & \cite{bian2014identifying} \\
\cellcolor{gray!25}International Conference on Electrical Engineering/Electronics, Computer & \cellcolor{gray!25}\cite{meananeatra2011using} \\
\cellcolor{gray!25}Telecommunications and Information Technology & \cellcolor{gray!25} \\
International conference on Aspect-oriented software development & \cite{ettinger2004untangling}\\
\cellcolor{gray!25}Conference on software engineering and advanced applications & \cellcolor{gray!25}\cite{charalampidou2018structural} \\
Annual Computer Software and Applications Conference & \cite{chen2017tool}\\
\cellcolor{gray!25}International Conference on Predictive Models and Data Analytics in Software Engineering & \cellcolor{gray!25}\cite{charalampidou2015size} \\
Transactions on Software Engineering and Methodology & \cite{vidal2018assessing} \\
\cellcolor{gray!25}International Conference on Software Maintenance & \cellcolor{gray!25}\cite{krishnan2013refactoring}\\
Conference on Software Maintenance, Reengineering, and Reverse Engineering & \cite{krishnan2014unification}\\
\cellcolor{gray!25}Software Engineering, Artificial Intelligence, Networking and Parallel/Distributed Computing & \cellcolor{gray!25}\cite{shin2019study} \\
International Workshop on Software Clones & \cite{goto2013extract,choi2011extracting} \\
\cellcolor{gray!25}Workshop on Software Evolution through Transformations & \cellcolor{gray!25}\cite{juillerat2006algorithm} \\
Symposium on Principles of Programming Languages & \cite{komondoor2000semantics}\\
\cellcolor{gray!25}ACM SIGPLAN workshop on Partial evaluation and program manipulation & \cellcolor{gray!25}\cite{brown2010clone,li2009clone}\\
Working Conference on Reverse Engineering & \cite{balazinska1999partial} \\
\cellcolor{gray!25}Seminar on Advanced Techniques Tools for Software Evolution & \cellcolor{gray!25}\cite{baars2019towards} \\

\bottomrule
\end{tabular}
\end{adjustbox}
\end{center}
\end{table}



\section{Results}
\label{Section:Result}
This section reports and discusses the results of our study.
\begin{table*}[ht!]
  \centering
	 \caption{\textcolor{black}{Related work in recommending the \textit{Extract Method} refactoring opportunities.}}
	 \label{Table:Related_Work_in_Extract_Method_Refatoring}

\begin{adjustbox}{width=1.0\textwidth,center}
\rowcolors{2}{white}{gray!25}
\begin{tabular}{lcllllllll}\hline
\toprule
\bfseries Study & \bfseries Year  &  \bfseries Intent & \bfseries Code Analysis & \bfseries Code Representation  & \bfseries Detection  & \bfseries Execution &  \bfseries Validation Method    \\
\midrule
 Lakhotia \& Deprez \cite{lakhotia1998restructuring} &  1998 &  Long Method & Semantic

 & Graphs  & Manual  & Suggest Alternatives& Proof of Concept \\ 

\textcolor{black}{Balazinska \etal \cite{balazinska1999partial}} & \textcolor{black}{1999} & \textcolor{black}{Code Clone} &  Syntactic & AST & Fully automated & Fully automated&   Proof of Concept\\

\textcolor{black}{Komondoor \& Horwitz \cite{komondoor2000semantics}} & \textcolor{black}{2000} & \textcolor{black}{Code Clone} & Semantic & Graphs  & Manual & Fully automated &  Proof of Concept \\

Maruyama~\cite{maruyama2001automated} & 2001 & 
Separation of Concerns 
& Semantic & Graphs & Manual & Choose Candidates &  Proof of Concept\\

\textcolor{black}{Komondoor \& Horwitz \cite{komondoor2003effective}} & \textcolor{black}{2003} & \textcolor{black}{Code Clone} &  Semantic & Graphs  & Manual  & Fully automated & Proof of Concept\\

\textcolor{black}{Ettinger \& Verbaere} \cite{ettinger2004untangling} & \textcolor{black}{2004} & \textcolor{black}{Separation of Concerns} & Semantic & Graphs & Manual  & Fully automated&  Proof of Concept \\

 Higo \etal \cite{higo2004refactoring} &  2004 & 
 Code Clone & Lexical & Tokens  & Fully automated &  Choose Candidates&  Case Study \\

 Higo \etal \cite{higo2004aries} &  2004  & 
 Code Clone & Semantic & Graphs & Fully automated & Fully automated & Case Study  \\

  Higo \etal \cite{higo2005aries} &  2005 & 
 Code Clone & Lexical & Tokens & Fully automated &  Execute on Approval &  Case Study  \\

  Higo \etal \cite{higo2008metric} &   2008 & 
 Code Clone & Textual & Source Code & Fully automated &  Execute on Approval&  Case Study \\

O'Connor \etal \cite{o2005star} & 2005 &
Separation of Concerns & Syntactic & AST & Semi-automated&  Suggest Alternatives &  Proof of Concept\\

\textcolor{black}{Juillerat \& Hirsbrunner} \cite{juillerat2006algorithm} & \textcolor{black}{2006} & \textcolor{black}{Code Clone} & Syntactic & AST & Fully automated & Fully automated& Proof of Concept  \\

\textcolor{black}{Juillerat \& Hirsbrunner} \cite{juillerat2007improving} & \textcolor{black}{2007} & \textcolor{black}{Separation of Concerns} & Syntactic & AST & Manual & Fully automated & Proof of Concept\\

 Vittek \etal \cite{vittek2007c++} &  2007 &
 Separation of Concerns
& Syntactic  & AST  & Manual & User Input & Proof of Concept \\

Corbat \etal \cite{corbat2007ruby} & 2007 
& Separation of Concerns
& Syntactic & AST & Manual & Choose Candidates  & Proof of Concept\\

 Murphy-Hill \& Black~\cite{murphy2008breaking} &  2008 &
 Separation of Concerns
&  Textual  &  Source Code & Manual  & Choose Candidates  & Experiment \\ 

\textcolor{black}{Abadi \etal} \cite{abadi2008re} & \textcolor{black}{2008} &
\textcolor{black}{Separation of Concerns}
& Textual  & Source Code  & Manual & Fully automated &  Case Study \\

Abadi \etal \cite{abadi2009fine} & 2009 &
Separation of Concerns
& Textual & Source Code & Manual & Fully automated &  Case Study\\

 Tsantalis \& Chatzigeorgiou~\cite{tsantalis2009identification} &  2009  &  
 Long Method &  Textual & Source Code  & Fully automated  &  Suggest Alternatives  & Experiment \\

 Tsantalis \& Chatzigeorgiou~\cite{tsantalis2011identification} &  2011 &
 Long Method&  Textual & Source Code  & Fully automated  &  Suggest Alternatives  & Experiment \\

Yang~\etal~\cite{yang2009identifying} & 2009 & Long Method
& Textual  & Source Code & Manual &  Suggest Alternatives & Case Study  \\

\textcolor{black}{Li \& Thompson} \cite{li2009clone} & \textcolor{black}{2009} & \textcolor{black}{Code Clone} & Hybrids & AST \& Tokens & Manual & Suggest Alternatives  & Case Study  \\

\textcolor{black}{Brown \& Thompson} \cite{brown2010clone} & \textcolor{black}{2010} & \textcolor{black}{Code Clone} &   Hybrids & AST \& Tokens & Manual &  Suggest Alternatives  & Case Study  \\

 Kanemitsu~\etal~\cite{kanemitsu2011visualization} &  2011 
&  Separation of Concerns
& Semantic  & Graphs  & Manual &  Suggest Alternatives & Experiment \\

Meananeatra \etal \cite{meananeatra2011using} & 2011 
& Long Method
& Syntactic & Metrics & Manual &Suggest Alternatives & Proof of Concept \\

\textcolor{black}{Choi \etal} \cite{choi2011extracting}
& \textcolor{black}{2011} & Code Clone & Lexical & Tokens & Fully automated & Manual & Case Study\\


 Sharma~\cite{sharma2012identifying} &  2012 &
 Separation of Concerns & Semantic
 & Graphs  & Manual  & Fully automated & Proof of Concept  \\

Cousot \etal \cite{cousot2012abstract} & 2012 & 
Separation of Concerns & Textual & Source Code & Manual & Fully automated & Proof of Concept \\ 

 Tairas \& Gray \cite{tairas2012increasing} &  2012 &
 Code Clone & Syntactic & AST & Fully automated  &  Choose Candidates &  Experiment  \\


Kaya \& Fawcett \cite{kaya2013identifying} & 2013 & 
Long Method & Textual & Source Code & Fully automated & Manual & Experiment \\ 

\textcolor{black}{Goto \etal \cite{goto2013extract}} & \textcolor{black}{2013} & \textcolor{black}{Code Clone} & Syntactic & AST & Manual & Fully automated & Case Study \\

 Bian \etal \cite{bian2013spape} &  2013 
&  Code Clone
&
  Hybrids & AST \& Graphs  & Manual & Fully automated &  Experiment  \\

   Bian \etal \cite{bian2014identifying} &  2014 
&  Code Clone
& Syntactic & Metrics  & Fully automated & Manual &  Experiment  \\

\textcolor{black}{Krishnan \& Tsantalis} \cite{krishnan2013refactoring} & \textcolor{black}{2013} & \textcolor{black}{Code Clone} & Textual  & Source Code & Fully automated &  User Input  & Experiment \\

\textcolor{black}{Krishnan \& Tsantalis} \cite{krishnan2014unification} & \textcolor{black}{2014} & \textcolor{black}{Code Clone} & Hybrids  & AST \& Graphs & Fully automated &  User Input  & Experiment \\

\textcolor{black}{Tsantalis \etal} \cite{tsantalis2015assessing} & \textcolor{black}{2015} & \textcolor{black}{Code Clone} & Hybrids & AST \& Source Code \& Tokens &  Fully automated &  User Input  & Experiment \\

\textcolor{black}{Mazinanian \etal} \cite{mazinanian2016jdeodorant} & \textcolor{black}{2016} & \textcolor{black}{Code Clone} & Hybrids & AST \& Source Code \& Tokens &  Fully automated &  User Input &  Experiment \\

\textcolor{black}{Tsantalis \etal} \cite{tsantalis2017clone} & \textcolor{black}{2017} & \textcolor{black}{Code Clone} & Hybrids & AST \& Source Code \& Tokens &  Fully automated &  User Input & Experiment \\

Silva~\etal~\cite{silva2014recommending} & 2014  &
Separation of Concerns
& Textual & Source Code & Fully automated &  Suggest Alternatives &  Experiment  \\

Silva~\etal~\cite{silva2015jextract} & 2015 &
Separation of Concerns
&  Textual & Source Code & Fully automated &  Suggest Alternatives &   Experiment    \\

 Fontana \etal \cite{arcelli2015duplicated} &  2015 &
 Code Clone & Hybrids 
 & AST \& Source Code  & Fully automated & Suggest Alternatives & Experiment \\ 

\textcolor{black}{Meng \etal} \cite{meng2015does} & \textcolor{black}{2015} & \textcolor{black}{Code Clone} & Syntactic & AST & Fully automated & Fully automated&  Experiment \\

\textcolor{black}{Charalampidou \etal \cite{charalampidou2015size}} & \textcolor{black}{2015} & \textcolor{black}{Long Method} &  Syntactic& Metrics & Fully automated  & Fully automated &  Case Study \\

Charalampidou~\etal~\cite{charalampidou2016identifying} & 2016 &
Long Method & Syntactic& AST \& Metrics & Fully automated  & Fully automated & Case Study \\

\textcolor{black}{Charalampidou \etal \cite{charalampidou2018structural}} & \textcolor{black}{2018} & \textcolor{black}{Long Method} &Syntactic& Metrics & Fully automated  & Fully automated &  Case Study \\

 Haas \& Hummel~\cite{haas2016deriving} &  2016  & 
 Long Method &
Hybrids  & Source Code \& Graphs  & Manual &  Suggest Alternatives & Experiment  \\

   Haas \& Hummel~\cite{haas2017learning} &  2017 & 
 Long Method &
Hybrids  & Source Code \& Graphs  & Manual &  Choose Candidates & Experiment  \\

Xu~\etal~\cite{xu2017gems} & 2017 & 
  Separation of Concerns & Textual
 & Source Code  & Fully automated &  Choose Candidates &Experiment  \\

 Imazato \etal \cite{imazato2017finding} &  2017 &  Separation of Concerns &  Textual &  Source Code  & Fully automated & Manual & Experiment \\

Kaya \& Fawcett \cite{kaya2017identification} & 2017 
& Long Method
& Semantic& Graphs  & Fully automated & Fully automated & Experiment \\ 

 Maruyama \& Hayashi \cite{maruyama2017tool} &  2017 
&  Separation of Concerns
&  Textual & Source Code  & Manual &  Choose Candidates &  Proof of Concept \\ 

Xu \etal \cite{xu2017log} & 2017 & Long Method  & Syntactic & Metrics & Fully automated & Manual &   Experiment \\

\textcolor{black}{Chen \etal} \cite{chen2017tool} & \textcolor{black}{2017} & \textcolor{black}{Code Clone} &  Syntactic  & AST & Manual & Fully automated & Case Study\\

\textcolor{black}{Ettinger \& Tyszberowicz} \cite{ettinger2016duplication} & \textcolor{black}{2016} & \textcolor{black}{Code Clone} & Textual  & Source Code &  Manual & Fully automated & Proof of Concept \\

\textcolor{black}{Ettinger \etal} \cite{ettinger2017efficient} & \textcolor{black}{2017} & \textcolor{black}{Code Clone} & Semantic & Graphs & Manual & Fully automated & Proof of Concept \\

 Meananeatra \etal \cite{meananeatra2018refactoring} &  2018 
&  Long Method
&  Hybrids & AST \& Graphs & Manual&  Execute on Approval &  Case Study  \\

Choi \etal \cite{choi2018investigation} & 2018 
& Long Method
& Syntactic & Metrics & Fully automated & Manual &  Experiment \\

 Yue~\etal~\cite{yue2018automatic} &  2018 
&   Code Clone & Syntactic 
  &  AST  & Fully automated  & Manual &  Experiment \\

\textcolor{black}{Vidal \etal} \cite{vidal2018assessing} & \textcolor{black}{2018} & \textcolor{black}{Long Method} & Textual  & Source Code & Fully automated&  Choose Candidates &Case Study\\

Yoshida~\etal~\cite{yoshida2019proactive} & 2019 
& Code Clone
& Hybrids & AST \& Tokens & Fully automated &  Choose Candidates &  Experiment \\

\textcolor{black}{Shin \cite{shin2019study}} & \textcolor{black}{2019} & \textcolor{black}{Code Clone} &  Syntactic & AST & Fully automated & Fully automated &  Case Study \\

\textcolor{black}{Barrs \& Oprescu \cite{baars2019towards}} & \textcolor{black}{2019} & \textcolor{black}{Code Clone} &  Hybrids & AST \& Graphs & Fully automated & Manual &  Experiment \\

 Antezana \cite{antezana2019toad} &  2019  & 
 Long Method & Textual
 &  Source Code  & Manual & Choose Candidates & Experiment  \\

 Alcocer \etal \cite{alcocer2020improving} &  2020 & 
 Long Method & Textual
 &  Source Code  & Manual & Choose Candidates &  Experiment  \\

Nyamawe \etal \cite{nyamawe2019automated} & 2019 
& Separation of Concerns
&  Textual & Text   & Fully automated & Manual &   Experiment \\

Nyamawe \etal \cite{nyamawe2020feature} & 2020 
& Separation of Concerns
&  Textual & Text   & Fully automated & Manual &  Experiment \\

 Krasniqi \& Cleland-Huang \cite{krasniqi2020enhancing} &  2020 &  
 Separation of Concerns &
 Textual & Text   & Fully automated & Manual &   Experiment \\

Abid \etal \cite{abid2020does} & 2020 
& Separation of Concerns
& Textual & Source Code & Manual  &  User Input & Experiment \\

 Sheneamer  \cite{sheneamer2020automatic} &  2020 &  Code Clone
&  Hybrids &  AST \& Graphs \& Tokens   & Fully automated  & Manual &  Experiment \\ 

Aniche~\etal~\cite{aniche2020effectiveness} & 2020 
& Separation of Concerns
&  Syntactic &  Metrics  & Fully automated & Manual &   Experiment \\

 Van~der~Leij~\etal~\cite{van2021data} &  2021 &  
 Separation of Concerns &
Syntactic &  Metrics  & Fully automated & Manual &  Experiment \\

Sagar \etal \cite{sagar2021comparing} & 2021 &
Separation of Concerns & Hybrids
 & Text \& Metrics  & Fully automated & Manual &  Experiment  \\

 AlOmar \etal \cite{alomar2022documentation} &  2022 & 
 Separation of Concerns & Textual &  Text  & Fully automated & Manual &  Experiment \\

Nyamawe \cite{nyamawe2022mining} & 2022 
& Separation of Concerns &
Textual &  Text  & Fully automated & Manual & Experiment \\

 Shahidi~\etal~\cite{shahidi2022automated} &  2022 &  
 Long Method &
Hybrids  & Graphs \& Metrics   & Fully automated & Fully automated &  Experiment \\

Tiwari \&  Joshi~\cite{tiwari2022identifying} & 2022 & Long Method & Semantic & Graphs & Fully automated & Manual &  Experiment \\

 Fernandes \etal \cite{fernandes2022liveref} &  2022 &  Long Method  
& Syntactic  &  Metrics & Fully automated&  Execute on Approval&   Experiment  \\

 Fernandes \etal \cite{fernandes2022live} &  2022 &  Long Method  
& Syntactic  &  Metrics & Fully automated&  Execute on Approval&  Experiment \\

AlOmar \etal \cite{alomar2022anticopypaster} & 2022 & Code Clone 
& Syntactic  &  Metrics & Fully automated&  Execute on Approval&  Experiment \\

AlOmar \etal \cite{alomar2023just} &  2023 & Code Clone 
&  Syntactic  &  Metrics & Fully automated& Execute on Approval&  Experiment \\

 Cui \etal \cite{cui2023rems} &  2023 &
 Separation of Concerns
& Semantic  &   Graphs & Fully automated & Manual&  Experiment \\

\textcolor{black}{Thy \etal \cite{thy15adventure}} & \textcolor{black}{2023} & \textcolor{black}{Separation of Concerns} & Textual & Source Code &  Fully automated& Fully automated &   Case Study \\

\textcolor{black}{Palit \etal \cite{palit14automatic}} & \textcolor{black}{2023} & \textcolor{black}{Separation of Concerns} &  Semantic&  Graphs & Fully automated  & Manual&  Experiment\\

\bottomrule
\end{tabular}
\end{adjustbox}

\vspace{-.3cm}
\end{table*}

\subsection{\RQone}

\begin{figure}[t]
\centering 
\begin{tikzpicture}
\begin{scope}[scale=0.75]
\pie[rotate = 180,pos ={0,0},text=inside,outside under=38,no number]{26.5/Long Method\and26.5\%, 34.9/SoC\and34.9\%, 38.6/Code Clone\and38.6\%}
\end{scope}
\end{tikzpicture}
\caption{\textcolor{black}{Percentage of \textit{Extract Method} studies, clustered by intent.}}
\label{fig:categories_clustered_rq1_intent}
\end{figure}
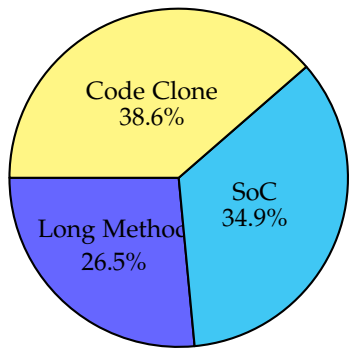

A detailed overview of the \textit{Extract Method} refactoring approaches reported by the \finalpool PSs is shown in Table~\ref{Table:Related_Work_in_Extract_Method_Refatoring}. Upon analyzing the PSs, we extract comprehensive high-level categories grouping the techniques used to implement the \textit{Extract Method} refactoring. These PSs are based on three main categories: (1) \textit{Code Clone},  \textit{Long Method}, and \textit{Separation of Concerns} (SoC). Figure~\ref{fig:categories_clustered_rq1_intent} shows the percentages of \textit{Extract Method} studies clustered by the detected intent. The \textit{Code Clone} category had the highest number of PSs, with a ratio of \textcolor{black}{38.6\%}. The \textit{Separation of Concerns} (SoC)  category accounted for \textcolor{black}{34.9\%}, with \textit{Long Method}  representing \textcolor{black}{26.5\%}. \textcolor{black}{Notably, these categories show minimal variation within the range of \textcolor{black}{26.5\%} to \textcolor{black}{38.6\%}. It should be noted that most of the \textit{Extract Method} refactoring tools (49\%) are primarily designed for the purpose of removing code clones.} In the rest of this section, we provide a more in-depth analysis of each of these categories along with the corresponding PSs.

\textbf{Category \#1: Code Clone.} This category refers to studies that are designed to recommend \textit{Extract Method} refactoring opportunities to eliminate \textit{Code Clone} design defects. Refactoring \textit{Code Clone} consists of taking a code fragment and moving it to create a new method while replacing all instances of that fragment with a call
to this newly created method. It is worth noting that some PSs \cite{komondoor2000semantics,balazinska1999partial,komondoor2003effective,higo2004aries,higo2005aries,higo2008metric,tairas2012increasing,bian2013spape,yue2018automatic,yoshida2019proactive,chen2017tool,krishnan2013refactoring,krishnan2014unification,tsantalis2015assessing,mazinanian2016jdeodorant,tsantalis2017clone,shin2019study,meng2015does,li2009clone,thompson2011haskell,higo2004refactoring,juillerat2006algorithm,brown2010clone,choi2011extracting,goto2013extract,bian2014identifying,ettinger2017efficient,ettinger2016duplication,baars2019towards,sheneamer2020automatic} utilized the concept of \textit{Code Clone} to consider some or all types of clones (\ie Type 1, Type 2, Type 3, Type 4), and others \cite{arcelli2015duplicated,alomar2022anticopypaster,alomar2023just} utilized \textit{Duplicate Code} by considering Type 1 clone.

\textcolor{black}{\textbf{Komondoor and Horwitz} \cite{komondoor2000semantics} proposed an algorithm to select statements that are worth extracting while ensuring semantics preservation. The authors identify conditions based on control and data dependencies, and the algorithm suggests moving the selected statements when the conditions hold}. \textcolor{black}{\textbf
{\texttt{CloRT}} \cite{balazinska1999partial} is developed to take into account the shared elements of cloned methods while utilizing the strategy design pattern to differentiate them. A dynamic pattern matching algorithm is used to identify the semantic distinctions between clones and their translation in terms of programming language entities. \textbf{Komondoor and Horwitz} \cite{komondoor2003effective} propose a semantic preserving algorithm for extracting difficult sets of statements, including the detection of duplicated fragments and extracting them into procedures,  to
make them extractable, achieving ideal results in more than 70\% of the difficult cases}. \textbf{\texttt{Aries}} \cite{higo2004aries,higo2005aries,higo2008metric} is an \textit{Extract Method} refactoring 
 tool based on code clone analysis on top of their previous tool \textbf{\texttt{CCShaper}} \cite{higo2004refactoring}, enabling users to select which clones to remove by characterizing code clones. \textcolor{black}{\textbf{Juillerat and Hirsbrunner} \cite{juillerat2006algorithm} propose an algorithm for \textit{Extract Method} refactoring to remove code clone. The algorithm first constructs the abstract syntax tree of Java code, then generates a list of tokens for clone identification, and finally identify clone that obeys certain constraints for \textit{Extract Method} refactoring}. \textcolor{black}{\texttt{\textbf{Wrangler}} \cite{li2009clone} is a hybrid approach based on tokens and AST
to detect code clones in Erlang/OTP programs automatically. The proposed clone detection approach is capable of reporting code
fragments that are syntactically identical and support clone removal using function extraction. \texttt{\textbf{HaRe}} \cite{brown2010clone} is designed for Haskell to detect and eliminate code duplication for function extraction.} \textcolor{black}{\textbf{Choi \etal} \cite{choi2011extracting} extract code clones for refactoring by combining clone metrics. Their observation is that the combinations of these metrics can identify refactorable clone classes with higher precision.} \textbf{\texttt{CeDAR}} \cite{tairas2012increasing} is an Eclipse plug-in that sends the results of clone detection data to Eclipse, and the IDE receives the information and determines which clones can be refactored by specifying the clones with specific properties to be refactored. This tool reportedly detects considerably more clone groups compared to open-source artifacts. \textcolor{black}{\textbf{\texttt{FTMPAT}} \cite{goto2013extract} introduces a method that relies on slice-based cohesion metrics to merge software clones. The method starts by taking two similar methods as input and first detect syntactic differences between them using AST differencing. Subsequently, it identifies pairs of code fragments within these methods,  to serve as suitable candidates for \textit{Extract Method}. Then, the identified candidates are then evaluated and prioritized using slice-based cohesion metrics}. \textbf{\texttt{SPAPE}} \cite{bian2013spape,bian2014identifying} is a near-miss clone extraction method applied to ten large-scale open-source software and reportedly can extract more clones than this software. \texttt{SPAPE} was initially developed in C programming language to refactor near-miss clones automatically. The tool utilizes a symbolic program execution to transform data and identify duplicated code to ensure cohesiveness for programmers. 

\textcolor{black}{\textbf{Krishnan \etal} \cite{krishnan2013refactoring,krishnan2014unification} propose an algorithm for refactoring of software clones with two objectives: 
 maximize the number of mapped statements and, at the same time, minimize the number of differences between the mapped statements. The authors compared the proposed technique with \texttt{CeDAR} and concluded that their approach can find
a significantly larger number of refactorable clones}. \textcolor{black}{In other studies \cite{tsantalis2015assessing,mazinanian2016jdeodorant,tsantalis2017clone}, \texttt{JDeodorant} has been extended to identify \textit{Extract Method} opportunities for \textit{Code Clone} extraction. The tool automatically assesses whether a pair of clones can be safely refactored while preserving the
behavior. The authors were able to increase the percentage of
refactorable clones to 36\% on the same clone dataset
used by Tairas and Gray \cite{tairas2012increasing}}.  Duplicated Code Refactoring Advisor (\textbf{\texttt{DCRA}}) \cite{arcelli2015duplicated} is released to select and suggest the best refactorings of duplicated code, aiming to reduce the human involvement during \textit{Duplicated Code} refactoring procedures.  The tool used NiCad \cite{roy2008nicad} for clone detection, which adds information characterizing every clone, \eg the clone's location in the class hierarchy, its size, and type. Next, through the refactoring advisor, the tool suggests the refactorings to remove the clones and provide a ranking of their quality.  \textcolor{black}{\textbf{RASE} \cite{meng2015does} is a clone removal tool that can apply combinations of six refactorings. \textit{Extract Method} is one of these refactroings used to extract common code guided by systematic edits}.   \textcolor{black}{\textbf{\texttt{PRI}} \cite{chen2017tool} employs refactoring pattern templates and traces cloned code fragments across
revisions. \texttt{PRI} takes as input the results from a clone
detector, and then automatically identifies refactored regions through
refactoring pattern rules in the subsequent revisions, and 
summarizes refactoring changes across revisions}.  \textcolor{black}{\textbf{Ettinger \etal} \cite{ettinger2017efficient,ettinger2016duplication} contribute to the automation of
type-3 clone elimination by preparation of non-contiguous code for extraction in a new method}. \textbf{\texttt{CREC}} \cite{yue2018automatic} is a learning-based approach that proposes specific clones through feature extraction. 
 The tool initially refactors R-clones (historically refactored) and NR-clones (typically not refactored). This process is carried out using 34 features that analyze the characteristics of each clone to classify them. The implementation of \texttt{CREC} is done in three stages: preparation of the clone data, training, and testing, which allows it to provide the programmer with an accurate refactoring recommendation. 
 
 \textbf{Yoshida \etal} \cite{yoshida2019proactive} released an \textit{Extract Method} refactoring tool to be used as a proactive clone recommendation system. The process is meant to be implemented as an Eclipse plug-in to keep track of changes in the code. This tool suggests changes in real-time versus at the end of the project. This routine makes the code fresh in the programmer's mind, allowing for more efficient progress. This is accomplished by actively tracking the user's work in Eclipse and suggesting edits. \textcolor{black}{\textbf{Shin} \cite{shin2019study} proposes a refactoring method for finding duplicate code used in branch statements and refactoring them by extracting common parts. The results of case studies with unskilled developers yielded an average of 10\% reduction in source code. \textbf{\texttt{CloneRefactor}} \cite{baars2019towards} detects code clones that are suitable for refactoring, based on their context and scope. Their results indicate that about 40\% of code duplication can be refactored by method extraction,
while other clones require other refactoring techniques}. \textbf{Sheneamer} \cite{sheneamer2020automatic} automatically extracts features from detected code clones and trains models to inform programmers of the type to refactor. Their approach categorizes refactored clones as distinct classes and develops a model to recognize the various types of refactored clones and those that are anonymous. \textbf{\texttt{AntiCopyPaster}} \cite{alomar2022anticopypaster,alomar2023just} is an IntelliJ IDEA plugin, implemented to detect and refactor duplicate code interactively as soon as a duplicate is created. The plugin only recommends the extraction of a duplicate only when it is \textit{worth it}, \ie the plugin treats whether a given duplicate code shall be extracted as a binary classification problem. This classification is performed using a CNN, trained using a dataset of 9,471 extract method refactorings of duplicate code collected from 13 open-source projects.
\begin{figure*}[t]
 	\centering
    \includegraphics[width=1.0\textwidth]{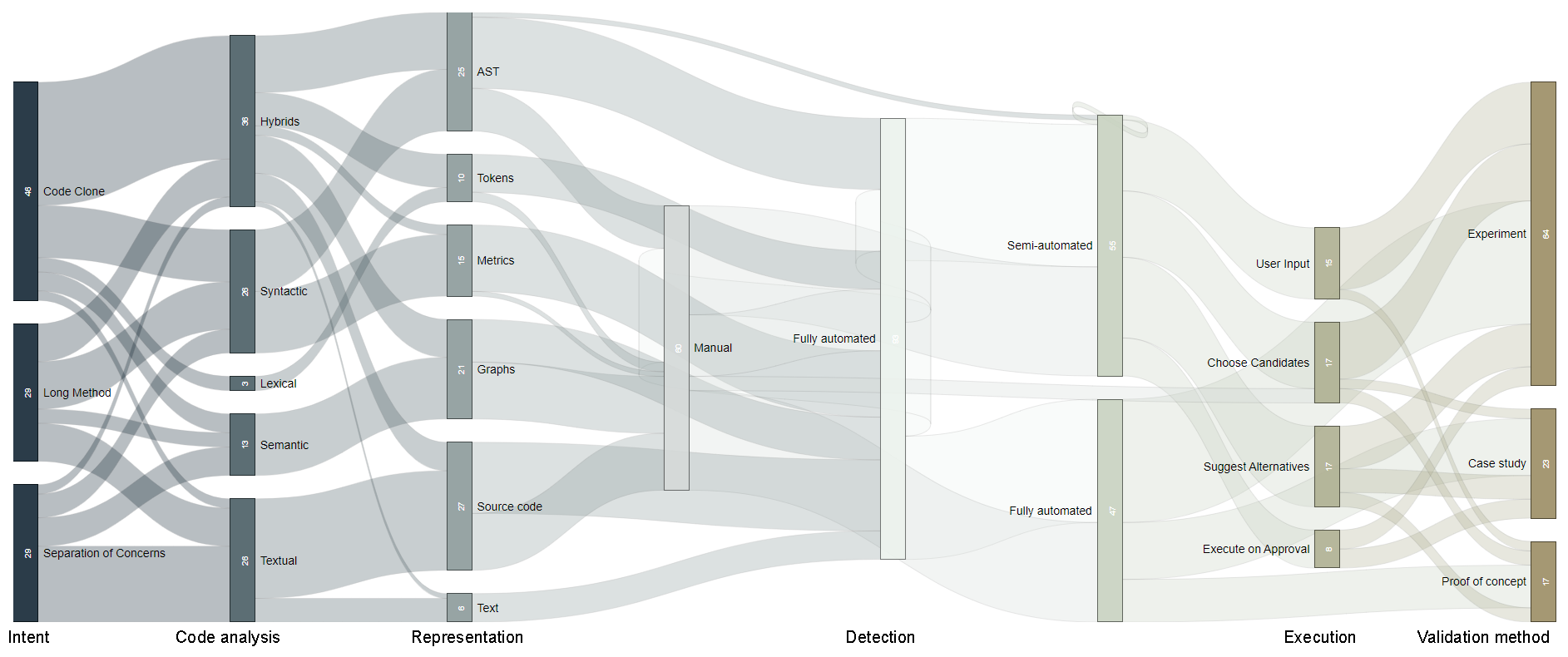}
 	\caption{\textcolor{black}{The relationship among the intent, code analysis, representation, detection, execution, and validation method of the \textit{Extract Method} refactoring.}} 
 	\label{fig:mapping}
\end{figure*} 

\textbf{Category \#2: Long Method.} This category refers \textcolor{black}{by} studies that are designed to identify \textit{Extract Method} refactoring opportunities to eliminate \textit{Long Method} design defects. \textit{Long Method} is a long and complex method that hinders the readability, reusability, and maintainability of the code. As a solution, refactoring \textit{Long Method} was proposed by extracting independent and cohesive fragments from long methods as new, short, and reusable methods \cite{tsantalis2009identification,tsantalis2011identification,yang2009identifying,charalampidou2016identifying,meananeatra2018refactoring,antezana2019toad,alcocer2020improving,tiwari2022identifying,fernandes2022liveref,fernandes2022live,lakhotia1998restructuring,xu2017log,vidal2018assessing,meananeatra2011using,kaya2013identifying,charalampidou2015size,charalampidou2018structural,haas2016deriving,haas2017learning,kaya2017identification,choi2018investigation,skiena1998algorithm,shahidi2022automated}.

Lakhotia and Deprez \cite{lakhotia1998restructuring} proposed a transformation tuck that restructures code and reorganizes unclear large fragments into small cohesive functions. \textbf{\texttt{Tuck}} \cite{lakhotia1998restructuring} deconstructs large functions into small functions by restructuring programs. Wedge, split, and fold are the three parts that makeup tuck. Then, statements of meaningful functions in a wedge are split and folded into a new function. \textbf{\texttt{JDeodorant}} \cite{tsantalis2009identification,tsantalis2011identification} encompassed identifying specific \textit{Extract Method} refactoring opportunities. This tool automatically identifies \textit{Extract Method} opportunities for \textit{Long Method} to suggest code improvement instead of requiring a set of statements from the programmer. Yang~\etal identified fragments to be extracted from long methods. Their approach is implemented as a prototype called \textbf{\texttt{AutoMed}} \cite{yang2009identifying}. The evaluation results suggested that the approach may reduce the refactoring cost by 40\%. \textbf{Meananeatra \etal} \cite{meananeatra2011using} proposed an approach to select refactorings dependent on data flow and control flow graphs of software metrics. The method procedure includes calculating metrics, filter refactorings, computing maintainability for candidate refactorings, then outlining \textit{Extract Method} refactorings 
   with the highest maintainability. The approach has been reported to accurately resolve \textit{Long Method} issues by suggesting refactoring techniques for the \textit{Extract Method}, replacing temp with the query, and decomposing condition. \textbf{Kaya and Fawcett} \cite{kaya2013identifying} automate selecting program refactoring fragments to resolve defects with the \textit{Long Method}. The paper goes over the identification process of code fragments based on a placement tree. This procedure outlines each node in the tree with variable reference counts to implement an effective process. \textcolor{black}{\textbf{Charalampidou \etal} \cite{charalampidou2015size,charalampidou2018structural} conduct a case study to evaluate several cohesion, coupling, and size metrics to serve as indicators of the existence of \textit{Long Method}, and
integrate these metrics into a multiple logistic regression
model, enabling the prediction of whether a method should be refactored or extracted}. The tool \textbf{\texttt{SEMI}} \cite{charalampidou2016identifying} ranks refactoring opportunities based on their extraction ability. This paper outlines \textit{Long Method}, to be implemented within a method to identify refactoring opportunities. The \texttt{SEMI} approach determines which parts of code are cohesive between statements. This can minimize the size of each method and create clear resulting methods that are increasingly single-responsibility principle compliant. This tool was validated with industrial and comparative case studies. 

\textbf{Hass and Hummel} \cite{haas2016deriving,haas2017learning} introduce refactoring and orders, each with a scoring function developed to reduce complexity and improve the way users read the code. This open-source software filters out invalid \textit{Extract Method} refactorings and then ranks to obtain different suggestions with the previously mentioned scoring function.  \textbf{Kaya and Fawcett} \cite{kaya2017identification} strive to implement \textit{Extract Method} refactoring and urge developers to utilize understandable implementation and modular structures so that the source code quality will not decrease throughout the project development. The goal is to refactor without requiring the user to select a code section. The approach searches for opportunities to refactor by declaring variables and regions of code that are fully extractable. The user can visualize the available refactoring options and choose which to apply without relying on a foreign code base. \textbf{\texttt{LLPM}} \cite{xu2017log} combines method-level software metrics applying a log-linear probabilistic model for accustomed refactorings. This application was tested with refactorings of real-world \textit{Extract Method} applications allowing the researchers to obtain parameter sets that capture the reason behind such refactorings. This analysis was completed by identifying the code to refactor and prioritizing various method groups to refactor. The proposed model optimizes parameters that maximize the probability of the collected dataset to refactor \textit{Long Method} bad smells accurately. \textbf{\texttt{LMR}} \cite{meananeatra2018refactoring} is an \textit{Extract Method} refactoring approach that utilizes program analysis and code metrics by implementing refactoring enabling conditions. This approach uses two guidelines for practical refactoring sets: code analyzability level and the statement number. Initially, \texttt{LMR} is applied to a Java application core package, showing that \textit{Long Method} bad smell can be eliminated in the code without removing behavior or making it more challenging to analyze. \textbf{Choi \etal} \cite{choi2018investigation} investigates change metrics and \textit{Extract Method} throughout two studies. The relationship results deduce a clear relationship between change metrics and \textit{Extract Method}. Product and change metrics must be available to accurately recommend refactorings for \textit{Extract Method}. The main contributions highlight metric change differences between extracted and not-extracted entities. \textcolor{black}{Vidal \etal \cite{vidal2018assessing} proposed \textbf{\texttt{Bandago}}, that is implemented on top of \texttt{JSpIRIT}, an Eclipse plugin for identifying and prioritizing code smells in Java. \texttt{Bandago} performs a heuristic search using a simulated annealing
algorithm \cite{skiena1998algorithm} that repeatedly applies the \textit{Extract Method} refactoring. Their findings reveal that the tool can automatically
fix more than 60\% of \textit{Brain Methods}, and when comparing the performance of \texttt{Bandago} with \texttt{JDeodorant}, the authors found that other types of code smells are also fixed after applying the \textit{Extract Method} refactoring suggestions}.  

\textbf{\texttt{TOAD}} \cite{antezana2019toad,alcocer2020improving} searches specific portions of the source code that include the developer's original code selection and meet ideal conditions for the \textit{Extract Method}. The approach operates during the workflow of refactorings and chooses fragments of code with correct syntax and outlined necessities. The tool explicitly recommends auto-refactoring alternatives when the user selects a piece of code and requests refactoring options. Overall, \texttt{TOAD}  reduced failed attempts significantly at a lower cognitive cost for \textit{Extract Method} refactoring. \textbf{Shahidi~\etal}~\cite{shahidi2022automated} automatically identified and refactored the \textit{Long Method} code smells in Java code using advanced graph analysis techniques. Their proposed approach was evaluated in five different Java projects. The findings reveal the applicability of the proposed method in establishing the single responsibility principle with a 21\% improvement. In another study, Tiwari and Joshi introduced \textbf{\texttt{Segmentation}} \cite{tiwari2022identifying} that identifies \textit{Extract Method} opportunities concentrating on achieving higher performance with fewer suggestions. Compared with other tools, \texttt{Segmentation} outperformed F-measure approaches and suggested that it showed high precision concerning small methods and \textit{Long Method} in opportunities with the \textit{Extract Method}. Empirical validations were applied to six open-source code applications to assess beneficial suggestions. \texttt{Segmentation} improves comparable recall and precision while identifying extract method refactorings. \textbf{\texttt{LiveRef}} \cite{fernandes2022live,fernandes2022liveref} is a tool implemented for live refactoring Java code. It works to resolve problems with long feedback loops that allow code to be maintainable and readable. The environment provides efficient refactoring suggestions by diminishing the time needed to apply, recommend, and identify the refactoring loop. The plugin for Java IntelliJ IDEA implemented a live refactoring environment that automatically applies \textit{Extract Method}. The tool results in improvements in the quality of the code along with faster programming solutions. 

\textbf{Category \#3: Separation of Concerns.} 
 The Separation of Concerns (SoC) category refers to studies segregating methods into multiple sub-methods based on their behavior so the code becomes less complex and effectively reused \cite{daga2006separation}. One of the main limitations of these studies \cite{o2005star,vittek2007c++,corbat2007ruby,murphy2008breaking,kanemitsu2011visualization,cousot2012abstract,silva2014recommending,silva2015jextract,xu2017gems,maruyama2017tool,ettinger2004untangling,thy15adventure,maruyama2001automated,juillerat2007improving,vittek2007c++,abadi2009fine,sharma2012identifying,imazato2017finding,nyamawe2019automated,nyamawe2020feature,krasniqi2020enhancing,abid2020does,aniche2020effectiveness,van2021data,sagar2021comparing,alomar2022documentation,nyamawe2022mining,cui2023rems,palit14automatic} is the absence of any context related to the application of refactorings, \ie it is not clear how developers would identify the need to apply these refactoring, \eg improving design metrics or removing design defects. \textbf{Maruyama} \cite{maruyama2001automated} solves the burden of manual refactoring by implementing automatic support when initiated by the programmer. It can be used by (1) selecting a fragment of code, (2) choosing a method, and (3) naming it. A new method is created from the parts of code from an existing method through block-based slicing. This mechanism is based on data-flow and control-flow analysis, so the user will not have to test the refactored fragment.   \textcolor{black}{\textbf{\texttt{Nate}} \cite{ettinger2004untangling} performs the \textit{ Extract Method}
refactoring by extracting the slice into a new method,
replacing it with a method call. For each extracted
statement, the tool determines whether to remove it from
the original method or to keep it there because it is still
relevant}. 
 \textbf{\texttt{SDAR}} \cite{o2005star} is an Eclipse plug-in that detects and applies local and global refactoring through star diagrams. The tool offers \textit{Extract Method} refactoring options that improve code and aid development opportunities and allows the refactoring
option for every node in the diagram that passes the JDT \textit{Extract Method} conditions.  
  \textcolor{black}{\textbf{Juillerat and Hirsbrunner} \cite{juillerat2007improving} construct an algorithm to recognize the arguments and outcomes of an extraction method. The implementation is an Eclipse plugin and uses the Java
Development Tools library provided by Eclipse.} 

 \textbf{\texttt{Xrefactory}} \cite{vittek2007c++} allows the application of \textit{Extract Method} refactoring using a back-mapping preprocessor to perform at the level of compilers in addition to other refactorings such as renaming, adding, and moving method parameters. 
  Although this tool only involves limited refactoring, the quality of the analysis indicates the quality of the whole refactoring tool. 
   \textbf{Corbat \etal} \cite{corbat2007ruby} developed a plug-in for the Eclipse Ruby development tools IDE since automated refactorings are not included in Ruby. Dynamic typing of Ruby makes implementing refactorings very difficult since it can be impossible for an IDE to determine an object type; therefore, \textit{Extract Method} refactoring was applied loosely adapted from JDT. 
    The tool \textbf{\texttt{RefactoringAnnotation}} \cite{murphy2008breaking} for \textit{Extract Method} refactoring allows the user to find solutions to coding errors. The annotations depend on what code section the programmer suggests and applies relevant refactoring recommendations. This is done automatically by implementing an arrow to be drawn on parameters and return values. The study concluded that speed, accuracy, and user satisfaction increase with the application of new tools. Usability recommendations are implemented, and the goal is to cultivate a new generation of tools that are user-friendly for programmers. \textcolor{black}{\textbf{Abadi \etal} \cite{abadi2008re} re-approach the
refactoring Rubicon by providing more general support for
method extraction. The authors performed a case study to convert a Java
servlet to use the model-view-controller pattern.} \textbf{Abadi \etal} \cite{abadi2009fine} introduces the foundation of fine slicing, a method that computes program slices. These slices can be transformed with the data removal and control dependencies as their surrounding code is extractable/executable.    
   \textbf{Cousot \etal} \cite{cousot2012abstract} highlight the problem of automatically inferring contracts such as validity, safety, completeness, and generality with method extraction. The proposed solution was to create two fast and capable tools that interact in an environment while maintaining precision. The practical solution is comprised of forward/backward methods that are iterative.   Silva~\etal~\cite{silva2014recommending} used a similarity-based approach to recommend automated \textit{Extract Method} refactoring opportunities that hide structural dependencies rarely used by the remaining statements in the original method. Their evaluation on a sample of 81 \textit{Extract
Method} opportunities achieved precision and recall rates close to 50\% when detecting refactoring instances. In another study, Silva~\etal~\cite{silva2015jextract} extended their work by designing an Eclipse plugin called \textbf{\texttt{JExtract}} that automatically identified, ranked, and applied refactorings upon request. The tool begins by generating all possibilities of \textit{Extract Method} for each method and then ranks these methods between dependencies in the code. 

\textbf{\texttt{ReAF}}~\cite{kanemitsu2011visualization} is a prototype tool that handles all Java language grammar. Initially, the user inputs source files to form a software system that the tool will visualize and build a procedural PDG for every method in the input. The tool can only handle Java source code but can be developed to handle other languages.  \textbf{Sharma}~\cite{sharma2012identifying} propose \textit{Extract Method} candidates based on the data and the structure dependency graph. Their suggestions were obtained by eliminating the longest dependency edge in the graph.
  \texttt{\textbf{GEMS}} \cite{xu2017gems} is an \textit{Extract Method} refactoring recommender that extracts structural and functional features related to complexity, cohesion, and coupling. It then uses this information to identify code fragments from a given source method that can be extracted. This method was tested comparatively with \texttt{JDeodorant} 
\cite{tsantalis2009identification,tsantalis2011identification}, \texttt{JExtract}  \cite{silva2014recommending,silva2015jextract} 
 and \texttt{SEMI} \cite{charalampidou2016identifying} 
   \textcolor{black}{to highlight the superiority of this tool.} The Eclipse plug-in was created to support software reliability with method extraction. \texttt{GEMS} validates potential code for a method and assigns a \say{goodness} score to it and recommends refactoring \textcolor{black}{with} \textit{Extract Method}. \textbf{Imazato \etal} \cite{imazato2017finding} propose a technique to find refactoring opportunities in the code using machine learning. The history of software development was analyzed as the basis of this tool to automatically suggest \textit{Extract Method} refactoring in the latest source code. This technique utilizes machine learning to identify potential refactoring opportunities. It consists of two phases: learning and predicting. The learning phase involves analyzing the characteristics of past cases and criteria, while the predicting phase involves detecting the location of possible refactorings. This design has the advantage of reducing the risk of overlooking refactorings. \textbf{\texttt{PostponableRefactoring}} \cite{maruyama2017tool} tool checks the code’s conditions and reports each defined error. These normal, fatal, and recoverable errors alert users when to apply the refactoring. Each error is refactorable since code may be rewritten altogether, but knowing which segments need work proves useful to programmers, especially throughout large projects.  \textbf{Nyamawe \etal} \cite{nyamawe2019automated,nyamawe2020feature} recommended \textit{Extract Method} refactorings based on the history of previously requested features, applied refactoring, and information about code smells. This learning-based approach is evaluated using a set of open-source projects with an F-measure of 70\% to recommend refactorings.  \textbf{Krasniqi and Cleland-Huang} \cite{krasniqi2020enhancing} develop a model first to detect refactoring commit messages from non-refactoring commits, then differentiate between 12 refactoring types. Their findings showed that SVM has an F-measure of 15\% when predicting \textit{Extract Method} refactorings.  \textbf{Abid \etal} \cite{abid2020does} highlights security throughout refactoring while attempting to improve various quality attributes. The proposed idea emphasizes security metrics and balancing code qualities through multi-objective refactoring. Compared with other approaches, this tool performs above existing approaches to improve the security of systems at a low cost while not sacrificing the quality of code. The paper determined that developers must prioritize security and other important qualities when establishing refactoring systems. 
  \textbf{Aniche~\etal}~\cite{aniche2020effectiveness} use a machine learning approach to predict refactorings using code, process, and ownership metrics. The resulting
models predict 20 different refactorings at the class, method, and variable levels. Their model achieved an accuracy of 84\% when predicting \textit{Extract Method} refactoring using Random Forest and Neural Network. Another experiment that predicts refactorings was conducted using quality metrics.  

\textbf{Van~der~Leij~\etal}~\cite{van2021data} explore the recommendation of the \textit{Extract Method} refactoring at ING. They observed that machine learning models could recommend \textit{Extract Method} refactorings with high accuracy, and the user study reveals that ING experts tend to agree with most of the model's recommendations. \textbf{Sagar \etal} \cite{sagar2021comparing} compare commit messages and source code metrics to predict \textit{Extract Method} refactoring. Their main findings show that the Random Forest
trained with commit messages or code metrics resulted in the best average accuracy of around 60\%. \textbf{AlOmar \etal} \cite{alomar2022documentation} formulate the prediction of refactorings as a multiclass classification problem, \ie classifying refactoring commits into six method-level refactoring operations, applying nine supervised machine learning algorithms. The prediction results for \textit{Extract Method} ranged from 63\% to 93\% in terms of F-measure. To predict \textit{Extract Method} refactorings, \textbf{Nyamawe} \cite{nyamawe2022mining} employs a binary classifier and recommends required refactorings with a multi-label classifier. This is done with the help of traditional refactoring detectors and commits message analysis to detect applied refactorings through machine learning.  
 \textbf{\texttt{REMS}} \cite{cui2023rems} recommend \textit{Extract Method}  refactoring opportunities via mining multi-view representations from code property graph. The results show that their approach outperforms four state-of-the-art refactoring tools, including \texttt{GEMS} \cite{xu2017gems}, \texttt{JExtract} \cite{silva2014recommending,silva2015jextract}, \texttt{SEMI} \cite{charalampidou2016identifying}, and \texttt{JDeodorant} \cite{tsantalis2009identification,tsantalis2011identification} in effectiveness and usefulness.  \textcolor{black}{\textbf{REM} \cite{thy15adventure} proposed an automated \textit{Extract Method}  built on
top of the IntelliJ IDEA plugin for Rust. Results reveal that  \texttt{REM} can extract a larger class of feature-rich code fragments into semantically correct
functions, can reproduce method extractions performed manually by
human developers, and is efficient enough to be used in interactive development}.  
\textcolor{black}{\textbf{Palit \etal} \cite{palit14automatic} employ a self-supervised autoencoder to acquire a representation of source code generated by a pre-trained large
language model for \textit{Extract Method} refactoring. Their experiments show
that their approach outperforms the state-of-the-art by 30\%
in terms of the F1 score.} 

\textcolor{black}{Next, we elaborate on the code analysis and code representation techniques as they were mentioned in their primary studies.}

\textbf{Code Analysis.}
\textcolor{black}{The nature of a code can be represented by the design properties of its specification. These properties can be decomposed into: (1) \textit{Textual}: no transformation or normalization is done to the source code, and generally the raw source code or textual information is used directly in the detection process; (2) \textit{Structural}: changes the source code into a series of lexical \say{tokens} using a compiler-style lexical analysis; (3) \textit{Syntactic}: employs a parser to transform source programs into parse trees or abstract syntax trees (ASTs). These can then be examined using either tree matching or structural metrics to detect code smells; 
 (4) \textit{Semantic}: captures the control and data flow of the program. It utilizes static program analysis to give more exact data than syntactic similarity. It generates a Program Dependence Graph (PDG), encompassing Control Flow Graphs (CFG) and Call Graphs (CG); and (5) \textit{Hybrids}: refers to techniques that use a combination of characteristics of other approaches.}

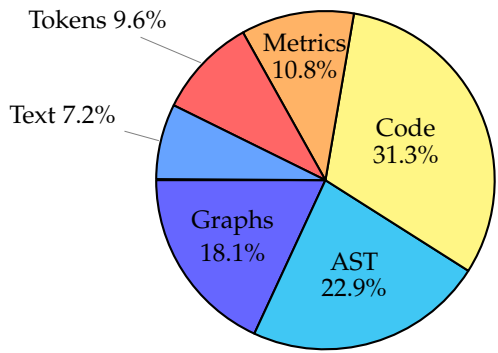
\begin{figure}[t]
\centering 
\begin{tikzpicture}
\begin{scope}[scale=0.75]
\pie[rotate = 180,pos ={0,0},text=inside,outside under=35,no number]{18.1/Graphs\and18.1\%, 22.9/AST\and22.9\%, 31. 3/Code\and31.3\%,10.8/Metrics\and10.8\%,9.6/Tokens\and9.6\%, 7.2/Text\and7.2\%}
\end{scope}
\end{tikzpicture}
\caption{\textcolor{black}{\textcolor{black}{Percentage of \textit{Extract Method} studies, clustered by code representation types.}}}
\label{fig:categories_clustered_rq1_representation}
\vspace{-.6cm}
\end{figure}

\textbf{Code Representation.}
\textcolor{black}{It spotlights the internal representation of the artifacts to be refactored. We extract comprehensive categories grouping the representation types used to implement the \textit{Extract Method} refactoring. These PSs are based on six main categories: (1) \textit{Source Code}, (2) \textit{Abstract Syntax Tree} (\textit{AST}), (3) \textit{Graphs}, (4) \textit{Metrics}, (5) \textit{Tokens}, and (6) \textit{Text}. Figure \ref{fig:categories_clustered_rq1_representation} illustrates the percentages of types of internal representation that the PSs used to make a decision on the extraction of the method. 
 As can be seen, 31.3\% of the PSs use \textit{Source Code} to recommend \textit{Extract Method} refactoring. Furthermore, 22.9\% of the approaches support the execution of the \textit{Extract Method} refactoring using \textit{AST}. The categories \textit{Graphs}, \textit{Metrics}, \textit{Tokens}, and \textit{Text} had the least number of PSs, with a ratio of  18.1\%, 10.8\%, 9.6\%, and 7.2\%, respectively.}

\textcolor{black}{We notice how the 3 \textit{Intent} clusters have used all categories of \textit{Code Analysis}, along with its associated types of \textit{Code Representation}. The \textit{Code Clone} cluster, despite being the largest in terms of studies, has the least number of papers that require developers to manually input the code to be refactored. This demonstrates how the existence of code clone detection tools has been supporting the refactoring studies since their early days. With the advancement in IDE support, studies shifted to automating the identification of refactoring opportunities, primarily by matching code smell patterns, then by mining patterns previously executed similar refactorings.}

\textcolor{black}{As for automating the recommendation, 53\% of the studies opted to include the developer in the loop. Incorporation can be in the form of asking for information to complete the transformation, such as requesting the name of the extracted method \cite{yamanaka2021recommending,bavota2014recommending}. 61\% of the studies provide multiple candidate solutions, either for the developer to choose from (\eg \cite{tairas2012increasing,haas2017learning}), or to also suggest other similar alternatives (\eg \cite{tsantalis2009identification,brown2010clone}).}

\textcolor{black}{For the \textit{Validation}, 16\% of mostly earlier studies handcrafted their own synthetic examples to assess the correctness of their solutions. The need for a more developer-centric assessment triggered validation to perform case studies. Evaluating the recommendation performance with developers provides a more grounded basis for judgement, at the expense of relatively specific setting that does not necessarily generalize. The rise of information retrieval in general, along with refactoring mining in particular, allowed studies to benefit from mined refactorings to assess accuracy and conduct comparative analysis.}

\textcolor{black}{Figure \ref{fig:mapping} provides detailed mappings between our six dimensions. We can observe that \textit{Code Clone} is the most popular intent-driving method extraction with a ratio of 38.6\%, followed up \textcolor{black}{by} \textit{Separation of Concerns}, taking 34.9\%, and finally \textit{Long Method} represented by 26.5\%. Interestingly, this is not matched in terms of the toolset, as the highest ratio of tools goes to \textit{Code Clone} with 49\%, then \textit{Long Method} and \textit{Separation of Concerns} with 26.5\% and 24.5\%, respectively. Such observation has caught our attention particularly as \textit{Separation of Concerns} is the only category that relies on all existing detection techniques and has its own unique one, \ie Evolutionary-based, and yet, there is a lack of concretizing this amount of research into practical tools. As for code representation, it is unsurprising that \textit{Code} is the most popular representation to identify need-to-refactor code fragments. This is being inherited from how research couples refactoring to a natural response to code smells, \eg \textit{Long Method}. So, metric-based detection rules are the most popular for detecting code smells \cite{sharma2018survey}, and so they become a go-to in the context of \textit{Extract Method}. Finally, existing studies offer a wide variety of static and dynamic techniques to execute the refactoring. They mainly rely on variants techniques of code slicing and graph analysis.} 


\begin{boxK}
\textit{\textbf{Summary.} \textcolor{black}{38.6\% of Extract Method refactoring studies are primarily addressing code clones. These studies commonly employ textual and structural code analysis as their internal representation to decide on method extraction. This representation is typically based on source code or Abstract Syntax Trees (AST).}} 
\end{boxK}

\subsection{\RQtwo}
\begin{table*}
  \centering
	 \caption{Characteristics of \textit{Extract Method} refactoring tools.}
	 \label{Table:Tool_Characteristics}

\begin{adjustbox}{width=1.0\textwidth,center}
\rowcolors{2}{white}{gray!25}
\begin{tabular}{lllllllllll}\hline
\toprule
\bfseries Tool & \bfseries Language      & \bfseries No of Metric &  \bfseries Interface & \bfseries Usage Guide? & \bfseries Tool Link & \bfseries Last Update \\
\midrule

  \textbf{Tuck} \cite{lakhotia1998restructuring} &   Unknown &   Unknown &   Unknown &   No &   Unknown &   Unknown\\

  \textcolor{black}{\textbf{CloRT}\cite{balazinska1999partial}} & \textcolor{black}{Java} & \textcolor{black}{N/A} & \textcolor{black}{Unknown} & \textcolor{black}{No} & \textcolor{black}{Unknown} & \textcolor{black}{Unknown}  \\

\textcolor{black}{\textbf{Nate}} \cite{ettinger2004untangling} & \textcolor{black}{Java} & \textcolor{black}{Unknown} & \textcolor{black}{Eclipse} & \textcolor{black}{No} & \textcolor{black}{Unknown} & \textcolor{black}{Unknown} \\

   \textbf{CCShaper} \cite{higo2004refactoring} &    Java &     6 &    \textcolor{black}{Command line} &    No &    Unknown &     Unknown \\

   \textbf{Aries} \cite{higo2004aries,higo2005aries,higo2008metric} &    Java &     6 &    \textcolor{black}{GUI-based} &    No &    Unknown &     Unknown \\ 

  \textbf{SDAR} \cite{o2005star} &   Java &   N/A  &   Eclipse &   No &   Unknown&    Unknown \\ 

  \textcolor{black}{\textbf{Unnamed} \cite{juillerat2007improving}} & \textcolor{black}{Java} & \textcolor{black}{N/A} & \textcolor{black}{Eclipse} & \textcolor{black}{No} & \textcolor{black}{Unknown} & \textcolor{black}{Unknown} \\

   \textbf{Xrefactory} \cite{vittek2007c++} &    C++ &     N/A &    Unknown&    Yes &    \cite{Xrefactory} &    2007 \\ 


  \textbf{Unnamed} \cite{corbat2007ruby} &   Ruby &   N/A &   Eclipse &   Yes &   \cite{Unnamed3} &   2012 \\ 

   \textbf{RefactoringAnnotation} \cite{murphy2008breaking} &   Java &    Unknown  &    Eclipse &    No &    Unknown &     Unknown \\ 

  \textbf{JDeodorant}~\cite{tsantalis2009identification,tsantalis2011identification,krishnan2013refactoring,krishnan2014unification,tsantalis2015assessing,mazinanian2016jdeodorant,tsantalis2017clone} &   Java &    3 &  IntelliJ / Eclipse &   Yes &   \cite{Jdeodorant} &   2019 \\

   \multirow{1}{*}{\textbf{AutoMed}~\cite{yang2009identifying}} &    Java   &    10 &     Unknown&    No &     Unknown &     Unknown\\ 

       \textcolor{black}{\multirow{1}{*}{\textbf{Wrangler}~\cite{li2009clone}}} &    \textcolor{black}{Erlang/OTP}   &    \textcolor{black}{N/A} &     \textcolor{black}{GUI-based / Command line}&    \textcolor{black}{Yes} &     \textcolor{black}{\cite{Wrangler}} &     \textcolor{black}{2023}\\ 

      \textcolor{black}{\multirow{1}{*}{\textbf{HaRe}~\cite{brown2010clone}}} &    \textcolor{black}{Haskell 98}   &    \textcolor{black}{N/A} &     \textcolor{black}{GUI-based / Command line}&    \textcolor{black}{Yes} &     \textcolor{black}{\cite{HaRe}} &     \textcolor{black}{2017}\\ 

  \multirow{1}{*} {\textbf{ReAF}~\cite{kanemitsu2011visualization}} &    Java   &     Unknown &    Unknown &    No &    Unknown &   Unknown  \\ 

   \textbf{Unnamed} \cite{cousot2012abstract} &    C\# &    Unknown &    Visual Studio extension &    No &    Unknown&     Unknown \\ 

  \textbf{CeDAR} \cite{tairas2012increasing} &   Java &    2  &   Eclipse &   No &   Unknown &   Unknown  \\ 

\textcolor{black}{\textbf{FTMPAT} \cite{goto2013extract}} & \textcolor{black}{Java} & \textcolor{black}{3} & \textcolor{black}{Eclipse} & \textcolor{black}{No} & \textcolor{black}{Unknown} & \textcolor{black}{Unknown}  \\

   \textbf{SPAPE} \cite{bian2013spape} &    Procedural / Java &      Unknown &    Unknown &    No &    Unknown &     Unknown \\ 

  \textbf{JExtract}~\cite{silva2014recommending,silva2015jextract} &   Java &    Unknown 
 &   Eclipse &   Yes &   \cite{JExtract} &   2016 \\

   \multirow{1}{*} {\textbf{DCRA} \cite{arcelli2015duplicated}} &    Java &      1 &     Unknown &    No &    Unknown&     Unknown \\ 

\textcolor{black}{\textbf{RASE}} \cite{meng2015does} & \textcolor{black}{Java} & \textcolor{black}{N/A} & \textcolor{black}{Eclipse} & \textcolor{black}{Yes} & \textcolor{black}{\cite{RASE}} & \textcolor{black}{2015} \\

  \textbf{SEMI}~\cite{charalampidou2016identifying} &   Java&      5  &   \textcolor{black}{GUI-based / Command line} &   Yes&   \cite{SEMI}&   2016\\


  \textbf{GEMS}~\cite{xu2017gems} &   Java &     48  &  Eclipse &   Yes &   \cite{GEMS} &   2017\\ 

   \textbf{PostponableRefactoring} \cite{maruyama2017tool} &    Java &    N/A &    Eclipse &    Yes &     \cite{PostponableRefactoring} &    2018 \\ 

  \textbf{LLPM} \cite{xu2017log} &    Java &   4 &    Unknown &   No &   Unknown &   Unknown \\

\textcolor{black}{\textbf{PRI}} \cite{chen2017tool} & \textcolor{black}{Java} & \textcolor{black}{N/A} & \textcolor{black}{Eclipse} &\textcolor{black}{No} & \textcolor{black}{Unknown} & \textcolor{black}{Unknown} \\

   \textbf{LMR} \cite{meananeatra2018refactoring} &    Java  &    5  &    Eclipse &    No &     Unknown &     Unknown \\ 

  \multirow{1}{*}{\textbf{CREC}~\cite{yue2018automatic}} &   Java &   N/A   &   Eclipse &   Yes &   \cite{CREC} &   2018 \\ 

\textcolor{black}{\textbf{Bandago}} \cite{vidal2018assessing} & \textcolor{black}{Java} & \textcolor{black}{4}& \textcolor{black}{Eclipse} & \textcolor{black}{No}& \textcolor{black}{Unknown}& \textcolor{black}{Unknown}\\

   \multirow{1}{*} {\textbf{Unnamed}~\cite{yoshida2019proactive}} &    Java   &    N/A   &    Eclipse &    No &    \cite{Unnamed2} &    2019 \\ 

   \textcolor{black}{\textbf{Unnamed}} \cite{shin2019study} & \textcolor{black}{Java} & \textcolor{black}{N/A}& \textcolor{black}{Unknown} & \textcolor{black}{No}& \textcolor{black}{Unknown}& \textcolor{black}{Unknown}\\

   \textcolor{black}{\textbf{CloneRefactor} \cite{baars2019towards}} & \textcolor{black}{Java} & \textcolor{black}{N/A} &  \textcolor{black}{Command line} & \textcolor{black}{No} & \textcolor{black}{\cite{CloneRefactor}} & \textcolor{black}{2020} \\

  \multirow{1}{*} {\textbf{\textcolor{black}{TOAD}} \cite{antezana2019toad,alcocer2020improving}} &   Pharo &   N/A  &   \textcolor{black}{Pharo} &   Yes &   \cite{TOAD} &   2019\\ 
\textbf{Segmentation}~\cite{tiwari2022identifying} & Java  &  2  & Eclipse & No & \cite{Segmentation} & 2022 \\

  \multirow{1}{*}{\textbf{LiveRef} \cite{fernandes2022liveref,fernandes2022live}} &   Java  &    20 &   IntelliJ &   Yes &   \cite{LiveRef} &   2022 \\ 
\textbf{AntiCopyPaster} \cite{alomar2022anticopypaster,alomar2023just} & Java &  78  & IntelliJ & Yes & \cite{AntiCopyPaster} &  2023 \\

\textcolor{black}{\textbf{REM} \cite{thy15adventure}} & \textcolor{black}{Rust} & \textcolor{black}{N/A} & \textcolor{black}{IntelliJ} & \textcolor{black}{Yes} & \textcolor{black}{\cite{REM}} & \textcolor{black}{2023}  \\

\bottomrule
\end{tabular}
\end{adjustbox}

\vspace{-.3cm}
\end{table*}

To help select an appropriate \textit{Extract Method} refactoring tool, we report in Table \ref{Table:Tool_Characteristics} the following main characteristics that can be considered to make an informed decision about tools usage:
\begin{itemize}
    \item \textit{Language}: Indicates the programming language the tool supports.
    \item \textit{Number of Metric}: Indicates the number of software metrics used by the tool.
    \item \textit{Interface}:  Indicates what IDE/user interface the tool supports.
    \item \textit{Usage Guide?}: Indicates the availability of instructions on how to use the tool.
    \item \textit{Tool Link}: \textcolor{black}{Points to} the online source code repository.
    \item \textit{Last Update}: Indicates whether the tool has been consistently updated/maintained since its development.
\end{itemize}

\begin{figure*}[t]
 	\centering
 	\includegraphics[width=1.0\textwidth]{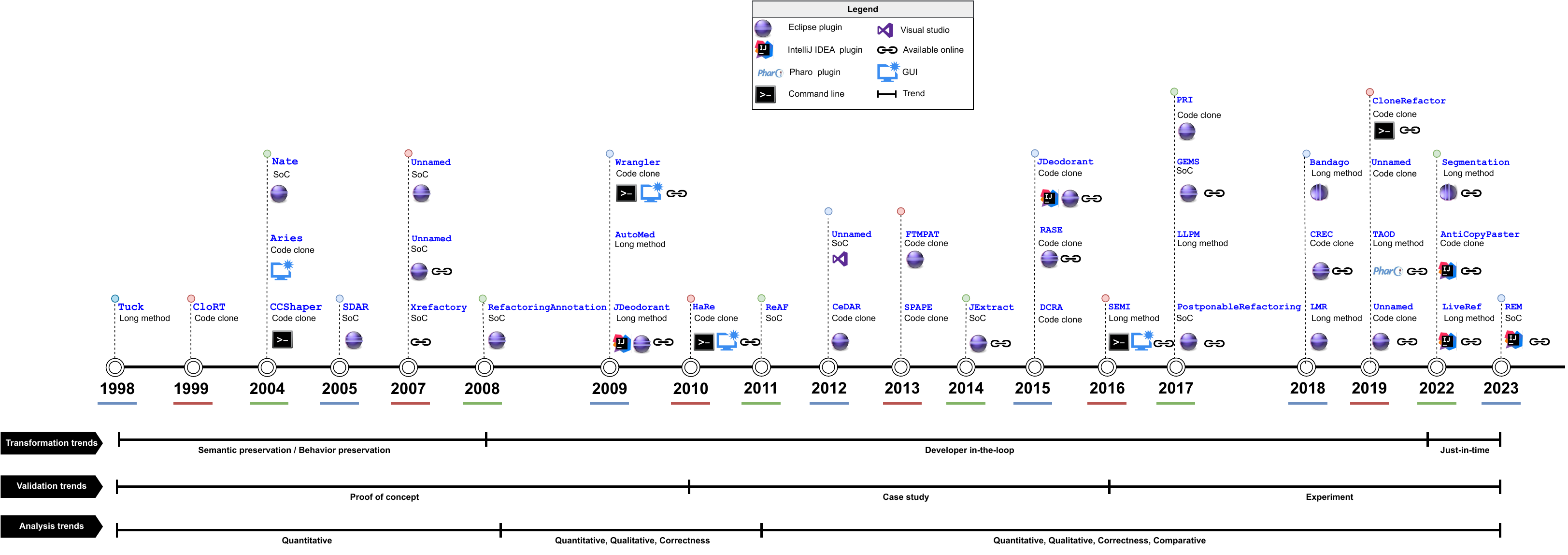}
 	\caption{\textcolor{black}{Timeline of developing \textit{Extract Method} refactoring tools.}}
 	\label{fig:timeline}
\end{figure*}

\textcolor{black}{Among the \finalpool primary studies, we identified \tool \textit{Extract Method} refactoring tools}. Table \ref{Table:Tool_Characteristics} provides the results for each of the \tool tools. We report any of these characteristics as `Unknown' in the table if we cannot locate the needed information and `N/A' if the information is not applicable to the study. It is evident from the table that the majority of \textit{Extract Method} tools are intended to recommend refactoring exclusively for Java-based systems. 
 As for metrics, most studies only mention quality attributes without the names of the metrics. Next, in terms of how developers interact with these tools, we found that most of the tools are in the form of IDE plugins, \ie Eclipse or IntelliJ, \textcolor{black}{and user interface or command line}. Regarding tool availability, we searched for a link to the tool website or binaries. In case the link is absent or no longer functional, we contacted the publication’s authors. From these \tool \textit{Extract Method} tools, we could only locate \toolonline tools. Figure \ref{fig:timeline} depicts a timeline of releasing \tool \textit{Extract Method} refactoring tools, in which \toolonline tools are made publicly available online by the research community. \textcolor{black}{There has been a considerable increase in the number of tools in the last two decades. The earlier tools were responsive to the challenge of ensuring the correctness of the transformation and its behavior preservation, given the lack of IDE support. The evaluation of these tools was mainly handcrafted, using fewer examples as a proof of concept. When IDEs started supporting the execution of code extraction, studies shifted toward automating the identification of refactoring opportunities while including developers in the tool design and evaluation. The rise of refactoring mining tools has enabled another dimension for studies to leverage previously performed extractions as \textit{ground truth} for predictive modeling, or for comparison baselines between existing solutions. Finally, recent techniques have taken a proactive fashion to immediately recommend refactoring, as soon as the opportunity is detected, in order to facilitate the adoption of the proposed change.}



\textcolor{black}{Several approaches have different automation support for detection and correction of \textit{Extract Method} refactoring identification. 
 In the rest of this section, we analyze the following level of automation for the \textit{Extract Method} refactoring tools.}

\textbf{Category \#1: Manual approach} \textcolor{black}{refers to using code inspection to detect or correct code smells.}

\textbf{Category \#2: Full automated approach} \textcolor{black}{refers to providing explicit full tool support to the users without human intervention.}

\textbf{Category \#3: Semi-automated approach}  \textcolor{black}{for the semi-automated approaches, it is broken down into four categories:}
\begin{itemize}
    \item \textcolor{black}{\textit{Suggest 
 Alternatives}: refers to the tool that is capable of carrying out the task automatically and proposing options or alternatives to the user. Nevertheless, the user must still manually select and implement the suggestion;}
    \item  \textcolor{black}{\textit{Choose Candidates}: refers to the tool that proposes alternative tasks to be done and requires the user to confirm the selection;}
    \item  \textcolor{black}{\textit{Execute on Approval}: refers to the tool that displays the activity that is about to be carried out and requests the user's permission. The user can either accept the activity in its entirety or cancel it;}
    \item \textcolor{black}{\textit{User Input}: refers to the tool that asks the user to select the code fragment as input to the tool.}
\end{itemize}
\textcolor{black}{Regarding the automaticity in the \textit{Extract Method} refactoring, we observe that most tools perform fully automated or semi-automatic refactoring tools. For example, the tool suggests an \textit{Extract Method}  refactoring for the code clone fragments, and the developer decides whether to apply or reject that refactoring. It is essential to highlight that automated refactoring alone cannot eliminate the need for manual verification after applying refactoring or manual refactoring in particular scenarios. That explains why many \textit{Extract Method} refactoring tools support semi-automatic refactoring. Furthermore, we observe that some tools utilize existing code smell detectors, and others integrate the detection of code smell and the execution of refactoring in the same tool. The latter eliminates the need to set up the dependency on a separate \textit{Long Method} splitter or \textit{Code Clone} detector.}

\begin{figure}[htbp]
 	\centering
 	\includegraphics[width=0.83\columnwidth]{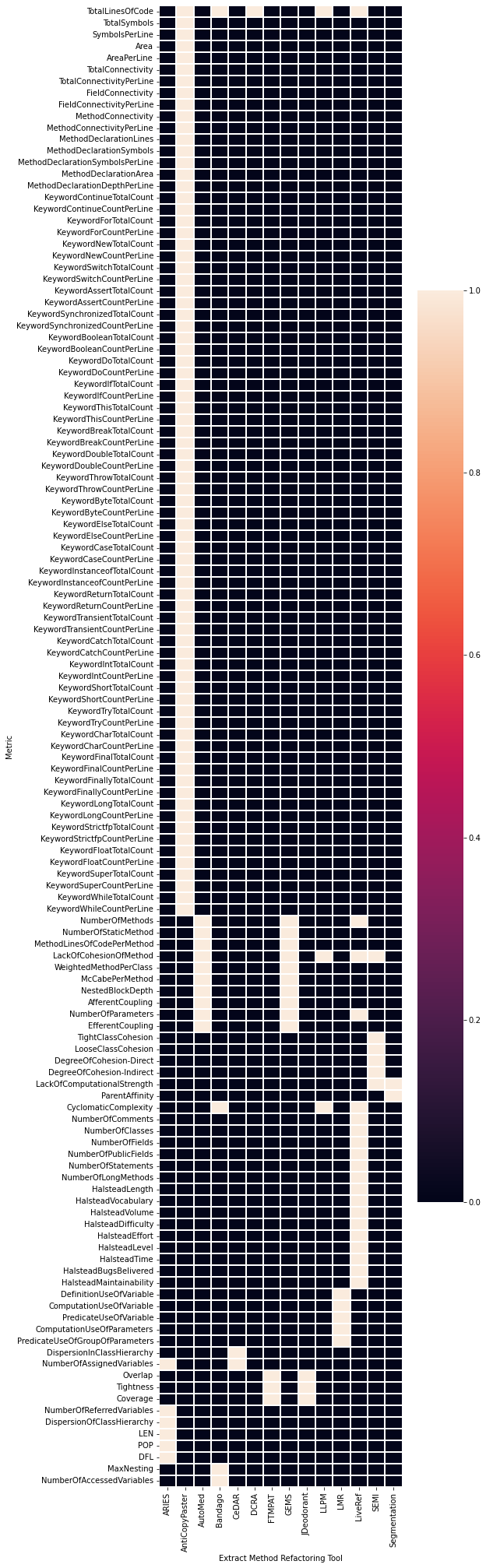}
 	\caption{\textcolor{black}{Software metrics considered in the \textit{Extract Method} refactoring tool.}}
 	\label{fig:metric}
\end{figure}

Figure \ref{fig:metric} depicts the software metrics used by the \metric \textit{Extract Method} refactoring tools (\textcolor{black}{the white color indicates that the tool computes the respective metric, while black signifies that the tool does not}). It is worth noting that we only include metrics that the PSs report. Some PSs indicated the usage of metrics without specifying the metric names. As can be seen, \metric of the \textit{Extract Method} refactoring tools, namely, \texttt{Aries}, \texttt{AntiCopyPaster}, \texttt{AutoMed}, \texttt{Bandago}, \texttt{CeDAR}, \texttt{DCRA}, \texttt{FTMPAT}, \texttt{GEMS}, \texttt{JDeodorant}, \texttt{LLPM}, \texttt{LMR}, \texttt{LiveRef}, \texttt{SEMI}, and \texttt{Segmentation}, indicated the metrics. These metrics relate to cohesion, coupling, complexity, size, keyword, and clone pairs. We found that `TotalLinesOfCode', `CyclomaticComplexity', `LackOfCohesionOfMethod', `NumberOfMethods', `NumberOfParameters', and `NumberOfAssignedVariables' are common metrics utilized by most of the tools. It should be noted that some of these metrics are used to assess quality improvement in refactoring research \cite{mohan2017multirefactor,alomar2019impact}. 
\begin{table*}
  \centering
	 \caption{Quantitative, qualitative, and comparative analysis of \textit{Extract Method} refactoring tools.}
	 \label{Table:Analysis}

\begin{adjustbox}{width=1.0\textwidth,center}
\rowcolors{2}{white}{gray!25}
\begin{tabular}{lllllllllll}\hline
\toprule
\bfseries Tool & \bfseries Quantitative     & \bfseries Qualitative & \bfseries Comparative & \bfseries Correctness \\
\midrule

    \textbf{Tuck} \cite{lakhotia1998restructuring} &     Unknown &    No &     No &     Unknown \\ 

      \textcolor{black}{\textbf{CloRT}\cite{balazinska1999partial}} & \textcolor{black}{Unknown} & \textcolor{black}{Unknown} & \textcolor{black}{Unknown}  & \textcolor{black}{Unknown} \\

\textcolor{black}{\textbf{Nate}} \cite{ettinger2004untangling} & \textcolor{black}{Unknown} &\textcolor{black}{No} & \textcolor{black}{No} & \textcolor{black}{Unknown} \\ 

    \textbf{CCShaper} \cite{higo2004refactoring} &     1 project &    No&     No &     Unknown \\ 

\textbf{Aries} \cite{higo2004aries,higo2005aries,higo2008metric} & 1 project &No& No & Unknown \\ 

    \textbf{SDAR} \cite{o2005star} &     Unknown &      No &     No &     Unknown\\ 

    \textbf{Xrefactory} \cite{vittek2007c++} & Unknown &No & No & Unknown\\ 


    \textbf{Unnamed} \cite{corbat2007ruby} &     Unknown &     No &     No &     Unknown\\ 

    \textbf{RefactoringAnnotation} \cite{murphy2008breaking} & 5 projects & w/ 16 developers & No & Unknown\\ 

    \textbf{JDeodorant}~\cite{tsantalis2009identification,tsantalis2011identification} &      1 project &     w/ 1 developer &     No &      Precision: 33.3\% - 100\%  \\
     &     &     &     &     Recall: 25\% - 100 \% \\
    &     &     &     &     Precision (AVG): 51\% \\
    &     &     &     &     Recall (AVG): 69\% \\ 
    \textbf{JDeodorant}~\cite{tsantalis2015assessing,mazinanian2016jdeodorant,tsantalis2017clone}     &      \textcolor{black}{9 projects} &     \textcolor{black}{No} &     \textcolor{black}{w/ CeDAR} &      \textcolor{black}{Accuracy: increase to 36\%}  \\
    \textbf{JDeodorant}~\cite{krishnan2013refactoring,krishnan2014unification}     &      \textcolor{black}{7 projects} &     \textcolor{black}{No} &     \textcolor{black}{w/ CeDAR} &      \textcolor{black}{Accuracy: increase to 83\%} \\ 

    \multirow{1}{*}{\textbf{AutoMed}~\cite{yang2009identifying}} &  1 project & No & No &     Accuracy: 3.57\% - 92.86\%  \\ 

    \textcolor{black}{\multirow{1}{*}{\textbf{Wrangler}~\cite{li2009clone}}}    &    \textcolor{black}{3 projects} &     \textcolor{black}{No}&    \textcolor{black}{No} &     \textcolor{black}{Unknown} \\

    \textcolor{black}{\multirow{1}{*}{\textbf{HaRe}~\cite{brown2010clone}}} &  \textcolor{black}{13 programs} & \textcolor{black}{No} & \textcolor{black}{No} &     \textcolor{black}{Unknown} \\ 

    \multirow{1}{*} {\textbf{ReAF}~\cite{kanemitsu2011visualization}} &     1 project &     w/ 14 developers &     w/ JDeodorant &     Unknown  \\ 

    \textbf{Unnamed} \cite{cousot2012abstract} & Unknown & w/ 4 authors & No & Unknown\\ 

    \textbf{CeDAR} \cite{tairas2012increasing} &     9 projects &    No&     w/ Aries \& Supremo* &     Unknown \\ 

\textcolor{black}{\textbf{FTMPAT} \cite{goto2013extract}} & \textcolor{black}{1 project} & \textcolor{black}{No} & \textcolor{black}{No} & \textcolor{black}{Unknown}  \\

    \textbf{SPAPE} \cite{bian2013spape} &     10 projects &     No&     No &     Unknown\\ 

    \textbf{JExtract}~\cite{silva2014recommending,silva2015jextract}  &     12 projects &     No &     w/ JDeodorant &      Precision: 38\% - 48\%
  \\ 
    &     &     &     &     Recall: 38\% - 48\%\\ 

    \multirow{1}{*} {\textbf{DCRA} \cite{arcelli2015duplicated}} &     50 projects &     No &     No &     Unknown \\ 

\textcolor{black}{\textbf{RASE}} \cite{meng2015does} & \textcolor{black}{2 projects} & \textcolor{black}{w/ experts} & \textcolor{black}{w/ RASE entire methods} & \textcolor{black}{Accuracy: 58\%}  \\

    \textbf{SEMI}~\cite{charalampidou2016identifying} &     5 projects &     w/ 3 developers &     w/ JDeodorant  &     Precision: 13.8\% - 22.4\% \\
    &     &     &     w/ JExtract &     Recall: 57.1\% - 92.8\%  \\
    &     &     &     &     F-measure: 22.23\% - 36.09\% \\ 


    \textbf{GEMS}~\cite{xu2017gems} &     5 projects &      w/ 4 authors &     w/ JDeodorant  &      Precision: 13.3\% - 25.3\% \\ 
    &     &     &     w/ JExtract&     Recall: 31.9\% - 49.2\% \\
    &     &     &     w/ SEMI &     F-measure: 18.8\% - 32.7\% \\ 

    \textbf{PostponableRefactoring} \cite{maruyama2017tool} &     Unknown &     No &     No &     Unknown \\ 

    \textbf{LLPM} \cite{xu2017log}  &     5 projects &     No &     w/ JDeodorant &     Precision: 18.5\% - 30.3\%  \\
    &     &     &      w/     JExtract &     Recall: 52.6\% - 62.1\% \\
    &     &     &     &     F-measure: 27.4\% - 40.7\% \\ 

    \textcolor{black}{\textbf{PRI}} \cite{chen2017tool} &     \textcolor{black}{6 projects} &     \textcolor{black}{No} &     \textcolor{black}{No} &     \textcolor{black}{Accuracy: 94.1\%}  \\

    \textbf{LMR} \cite{meananeatra2018refactoring} &     1 project &     No &    No &     Unknown \\ 

    \multirow{1}{*}{\textbf{CREC}~\cite{yue2018automatic}}  &     6 projects &     No &     No &    F-measure: 76\% - 83\%  \\ 

\textcolor{black}{    \textbf{Bandago}} \cite{vidal2018assessing} & \textcolor{black}{    10 projects} & \textcolor{black}{    w/ 35 developers} & \textcolor{black}{    w/ JDeodorant}& \textcolor{black}{    Unknown}\\

    \textcolor{black}{\textbf{Unnamed}} \cite{shin2019study} &     \textcolor{black}{Unknown} &     \textcolor{black}{w/ 6 teams}&     \textcolor{black}{No} &     \textcolor{black}{Unknown}\\

    \multirow{1}{*} {\textbf{Unnamed}~\cite{yoshida2019proactive}} &     2 projects &     w/ 8 developers &     No &     Unknown\\ 
    
 \textcolor{black}{\textbf{CloneRefactor} \cite{baars2019towards}} & \textcolor{black}{1,343 projects} & \textcolor{black}{No} & \textcolor{black}{No} & \textcolor{black}{Unknown}  \\
 
    \multirow{1}{*} {\textbf{\textcolor{black}{TOAD}} \cite{antezana2019toad,alcocer2020improving}} &     9 projects &     w/ 10 developers &     No &     Unknown \\ 

     \textbf{Segmentation}~\cite{tiwari2022identifying}   &     6 projects &     No&    w/ JExtract &     Precision: 22.81\% - 38.75\%  \\
    &     &     &     w/ SEMI &     Recall: 24.58\% - 41.75\% \\
    &     &     &     &     F-measure: 23.66\% - 40.19\% \\ 

    \textbf{LiveRef} \cite{fernandes2022liveref,fernandes2022live} &     3 projects &     w/ 42 developers &     No &     Unknown \\ 

    \textbf{AntiCopyPaster} \cite{alomar2022anticopypaster,alomar2023just} &     13 projects &     w/ 72 developers &     No &     Precision: 82\%  \\
    &     &     &     &     Recall: 82\% \\
    &     &     &     &     F-measure: 82\% \\
    &     &     &     &     PR-AUC: 86\% \\ 

    \textcolor{black}{\textbf{REM} \cite{thy15adventure}} &     \textcolor{black}{5 projects} &     \textcolor{black}{No} &     \textcolor{black}{w/ IntelliJ’s Rust} &     \textcolor{black}{Unknown}   \\

    &     &     &      \textcolor{black}{w/ Visual Studio
Rust Analyzer} &     \\
\bottomrule
\footnotesize{`$*$' indicates the tool is not peer-reviewed}
\end{tabular}
\end{adjustbox}

\vspace{-.3cm}
\end{table*}


 Table \ref{Table:Analysis} shows the quantitative, qualitative, comparative, and correctness data analysis of \textit{Extract Method} refactoring tools. \textcolor{black}{It is evident from the table that there is a noticeable absence of validation-related information from both quantitative and qualitative perspectives}. While the quantitative analysis seems to be the default experimentation by most of the primary studies, only 34\% reported the correctness of their tools through the standard performance metrics (\eg precision, recall). On the other hand, 26\% of tools were purely evaluated qualitatively. Only 15\% of the tools undergo both quantitative and qualitative analysis. Moreover, \texttt{JDeodorant} and \texttt{JExtract} are widely used by 23\% of the studies for comparative analysis.
To summarize, most studies rely on quantitative analysis or qualitative analysis to create oracles for their recommendation. Therefore, they need to go beyond the correctness and investigate the usefulness of their recommendations from the developer's standpoint, which was done only for 15\% of the tools. Additionally, many studies do not position their recommendations properly with respect to existing literature reviews through proper comparative analysis. Regarding correctness, most tools do not indicate details around their accuracy. From the set of \tool \textit{Extract Method} tools, only \toolaccuracy tools provide information about the tool's accuracy. 

\begin{boxK}
\textit{\textbf{Summary.} 
\textcolor{black}{A total of \tool Extract Method refactoring tools have been developed, with 49\% designed for refactoring code clones and 24\% intended to break down lengthy methods. Among these tools, approximately 58\% are developed as plugins, 9\% are command-line tools, and 9\% feature graphical user interfaces (GUIs). Several of these tools incorporate the developer's involvement in the decision-making process when applying the method extraction.}}
\end{boxK}

\subsection{\RQthree}
\textcolor{black}{
We investigate the datasets, and benchmarks that are used to evaluate and validate \textit{Extract Method} refactoring studies. We follow the same extraction procedure as described in Abgaz \etal \cite{abgaz2023decomposition}. A summary of the findings is illustrated in Tables \ref{Table:Tool_Benchmark-longmethod}, \ref{Table:Tool_Benchmark-codeclone}, and \ref{Table:Tool_Benchmark-SoC}.}


\textbf{Codebases.} \textcolor{black}{The evaluation of proposed \textit{Extract Method} studies depends on the availability of datasets and benchmarking data, which is a relatively unexplored area. We identified that most of the studies used a dataset created by the paper's authors, corresponding to 
 86.74\%. Only
 13.25\% reused datasets from previous studies. The selection of applications for experimentation is based on the availability of the source code, and the \textit{Extract Method} tools. Due to the absence of agreed-upon evaluation benchmarks, studies have generally used custom evaluations. Generally, PSs have mostly employed relatively small- or medium-scale open-source applications, typically containing less than 225,000 lines of code. Examples of open-source systems utilized by some PSs with the intent of \textit{Long Method} and \textit{Separation of Concerns} include JHotDraw and JUnit. Ant and JFreeChart are becoming popular Java systems for \textit{Extract Method} evaluation when extracting code clone\footnote{Due to space constraints, we report project names if the number of projects considered is less than or equal to 15.}.}

\textbf{Validation Methods.} \textcolor{black}{Various structured evaluation approaches have been suggested, such as proof of concepts, case studies, and experiments. Proof of concept involves demonstrating how the identification process works with the help of examples. Case studies examine the migration process in depth by looking at relevant cases, using one or multiple projects as a target. Experiments involve selecting the chosen codebases and then experimentally evaluating them using metrics such as coupling, cohesion, complexity, and code size, or comparing them with other tools. It should be noted that validation methods are reported as they were mentioned in their primary studies.}

\textcolor{black}{Previous studies have classified validation methods into \textit{proof of concepts}, \textit{case studies}, and \textit{experiments} \cite{abgaz2023decomposition,fritzsch2019microservices}. In our study, \textit{experiment-based} validation is the most widely used method, with 59.03\% of the studies that use it \cite{murphy2008breaking,tsantalis2009identification,tsantalis2011identification,kanemitsu2011visualization,tairas2012increasing,bian2013spape,silva2014recommending,silva2015jextract,arcelli2015duplicated,charalampidou2016identifying,xu2017gems,yue2018automatic,yoshida2019proactive,tiwari2022identifying,fernandes2022liveref,fernandes2022live,alomar2022anticopypaster,alomar2023just,krishnan2013refactoring,krishnan2014unification,tsantalis2015assessing,mazinanian2016jdeodorant,tsantalis2017clone,meng2015does,li2009clone,meananeatra2011using,kaya2017identification,haas2016deriving,haas2017learning,kaya2017identification,xu2017log,choi2018investigation,antezana2019toad,alcocer2020improving,shahidi2022automated,li2009clone,sheneamer2020automatic,imazato2017finding,nyamawe2019automated,nyamawe2020feature,krasniqi2020enhancing,abid2020does,aniche2020effectiveness,van2021data,sagar2021comparing,alomar2022documentation,nyamawe2022mining,cui2023rems,palit14automatic}. Some of these studies even combined a survey or user study with their experiment (\eg \cite{fernandes2022live,fernandes2022liveref,alomar2022anticopypaster,alomar2023just,vidal2018assessing}). The \textit{case study} is the second most dominant method, with 21.68\% of the papers applying it to evaluate their methods \cite{higo2004aries,higo2005aries,higo2008metric,yang2009identifying,meananeatra2018refactoring,vidal2018assessing,chen2017tool,shin2019study,goto2013extract,thy15adventure,thompson2011haskell,charalampidou2015size,charalampidou2018structural,brown2010clone,choi2011extracting,abadi2008re,abadi2009fine}. \textit{Proof of concept} method was also adopted by 19.27\% \cite{lakhotia1998restructuring,o2005star,vittek2007c++,corbat2007ruby,cousot2012abstract,ettinger2004untangling,balazinska1999partial,komondoor2000semantics,komondoor2003effective,juillerat2006algorithm,ettinger2016duplication,ettinger2017efficient,maruyama2001automated,juillerat2007improving,sharma2012identifying}. It is evident that experiment-based validation is becoming more popular. This is likely due to recent advances in metrics and benchmarks that make it easier to compare different \textit{Extract Method} techniques.}

\textbf{Programming Languages.} \textcolor{black}{The majority of studies (81.92\%) centralize on Java-based applications \cite{higo2004aries,higo2005aries,higo2008metric,o2005star,murphy2008breaking,tsantalis2009identification,tsantalis2011identification,yang2009identifying,kanemitsu2011visualization,tairas2012increasing,bian2013spape,silva2014recommending,silva2015jextract,arcelli2015duplicated,charalampidou2016identifying,xu2017gems,maruyama2017tool,xu2017log,meananeatra2018refactoring,yue2018automatic,yoshida2019proactive,tiwari2022identifying,fernandes2022liveref,fernandes2022live,alomar2022anticopypaster,alomar2023just,vidal2018assessing,chen2017tool,ettinger2004untangling,krishnan2013refactoring,krishnan2014unification,tsantalis2015assessing,mazinanian2016jdeodorant,tsantalis2017clone,shin2019study,goto2013extract,meng2015does,balazinska1999partial,meananeatra2011using,charalampidou2015size,charalampidou2018structural,haas2016deriving,haas2017learning,choi2018investigation,shahidi2022automated,juillerat2006algorithm,choi2011extracting,ettinger2016duplication,ettinger2017efficient,sheneamer2020automatic,maruyama2001automated,juillerat2007improving,abadi2008re,abadi2009fine,imazato2017finding,nyamawe2019automated,nyamawe2020feature,krasniqi2020enhancing,abid2020does,aniche2020effectiveness,van2021data,sagar2021comparing,alomar2022documentation,nyamawe2022mining,cui2023rems,palit14automatic}, while C++ \cite{vittek2007c++,bian2013spape,bian2014identifying}, Ruby \cite{corbat2007ruby}, C\# \cite{cousot2012abstract}, Pharo \cite{antezana2019toad,alcocer2020improving}, Haskell \cite{thompson2011haskell}, Erlang/OTP \cite{li2009clone} and Rust \cite{thy15adventure}, Java and Procedural in combination \cite{bian2013spape}, accounts for 18.07\%. It is evident that \textit{Extract Method} studies tend to incorporate Java codebases. This could be because many tools \textit{Extract Method} are designed for Java.}

\textbf{Dataset Availability.} \textcolor{black}{Dataset availability is one of the essential factors that allow the reproducibility and extension of studies. We collect all artifacts associated with the PSs, which encompasses studies providing raw datasets that require processing by researchers, as well as those that offer solely user survey responses from developers. It is observed from Tables \ref{Table:Tool_Benchmark-longmethod}, \ref{Table:Tool_Benchmark-codeclone}, and \ref{Table:Tool_Benchmark-SoC}  that 78.31\% of \textit{Extract Method} datasets are not publicly available. This observation highlights the need for public datasets to enable replication and extension of studies and mitigate benchmark bias when comparing the proposed approach with existing studies.}

We conjecture that the ground truth used to compare with existing studies might be biased. Also, the comparison against the state-of-the-art may not be appropriate unless these tools are called in the same context or intent as in the original paper. For instance, \texttt{JDeodorant} applies the \textit{Extract Method} refactoring to deal with long methods. If this tool is being tested against an \textit{Extract Method} performed to remove duplicates, it is expected not to recommend any code changes. Therefore, performing experimentation with techniques that address different intents may not be adequate. In a similar context, building a universal model that extracts methods based on the history of code changes without understanding the intent must be human-verified to see whether it is useful.

\begin{table*}
  \centering
	 \caption{\textcolor{black}{Benchmarks and datasets used in \textit{Extract Method} refactoring studies for \textit{Long Method} decomposition.}}
	 \label{Table:Tool_Benchmark-longmethod}

\begin{adjustbox}{width=1.0\textwidth,center}
\rowcolors{2}{white}{gray!25}
\begin{tabular}{lllllllllll}\hline
\toprule
\bfseries Study & \bfseries Intent & \bfseries Language      & \bfseries No of Metric &  \bfseries No of Project & \bfseries Project & \bfseries Other Properties & \bfseries Dataset Link & \bfseries Validation Method \\
\midrule

\textbf{Tuck} \cite{lakhotia1998restructuring} &     Long Method &    Unknown &    Unknown &     Unknown &    Unknown &    Unknown &    Unknown &    Proof of Concept\\

\textbf{JDeodorant}~\cite{tsantalis2009identification,tsantalis2011identification} & Long Method & Java &  3 &1 & Violet 0.16 & LOC: 4,100/ 61 classes/ 144 methods & Unknown & Experiment \\

   \multirow{1}{*}{\textbf{AutoMed}~\cite{yang2009identifying}} &    Long Method &     Java   &    10 &     1&    houtReader 1.8.0 &    LOC: ~20,000 / 269 classes &     Unknown &     Caee Study\\ 

\textbf{Meananeatra \etal} \cite{meananeatra2011using} 
& Long Method
& Java & 3 & \textcolor{black}{Unknown} & \textcolor{black}{Unknown} & Unknown & \textcolor{black}{Unknown} & \textcolor{black}{Experiment} \\ 

   \textbf{Kaya \& Fawcett} \cite{kaya2013identifying} 
&    Long Method &
   C++ &    N/A &    Unknown &    Unknown &    Unknown &    Unknown &    \textcolor{black}{Experiment} \\

\textcolor{black}{\textbf{Charalampidou \etal} \cite{charalampidou2015size}}  & \textcolor{black}{Long Method} & Java & 5 & 1 & jFlex & Unknown & Unknown &  Caee Study \\

   \textcolor{black}{\textbf{Charalampidou \etal} \cite{charalampidou2018structural}}  &    \textcolor{black}{Long Method} &    Java &    8 &    1 &    jFlex &    Unknown  &    Unknown &    Caee Study  \\

\textbf{SEMI}~\cite{charalampidou2016identifying} & Long Method & Java&    5  & 5 & Wikidev& Unknown& \cite{SEMI-dataset}& Caee Study\\
& & &&  &MyPlanner&   &   &   \\
  &   &   &  &    &  MyWebMarket&   &  &    \\
  &   &   &  &    &  JUnit&   &  &    \\
  &   &   && &JHotDraw & & & \\

    \textbf{Haas \& Hummel}~\cite{haas2016deriving} &  
    Long Method &    Java &    2 &    3 &    Agilefant &    LOC: 36,116/ 2,841 methods &    Unknown &    Experiment \\
   &    &    &    &    &    JabRef &    LOC: 128,145 / 5,665 methods &     &     \\
   &    &    &    &    &    JChart2D &    LOC: 50,728 / 1,849 methods &     &     \\

 \textbf{Haas \& Hummel}~\cite{haas2017learning} &  
 Long Method & Java & 9 & 13 & Unknown & Unknown & Unknown & Experiment \\

   \textbf{Kaya \& Fawcett} \cite{kaya2017identification} 
&    Long Method &
   C++ &    N/A &    Unknown &    Unknown &    Unknown &    Unknown &    \textcolor{black}{Experiment} \\

    \textbf{LLPM} \cite{xu2017log} &  Separation of Concerns &  Java &  4 &   5 & Wikidev &  130 total methods &  Unknown &  Experiment \\
 &  &  & &   & SelfPlanner&   & &   \\
 &  &  & &   & MyWebMarket&   &  &   \\
 &  &  & &   & JUnit&   & &   \\
 &  &  & &   & JHotDraw &   &  &   \\


\textbf{LMR} \cite{meananeatra2018refactoring} & Long Method & Java  & 5  & 1 & JFreeChart 1.0.17 & LOC: 5,665 / 20 classes / 552 methods&  Unknown &  Caee Study \\ 

   \textbf{Choi \etal} \cite{choi2018investigation} 
&    Long Method
&    Java &    6 &    1 &    JEdit &    LOC: 97,116 - 313,706 &    Unknown &    Experiment  \\

\textcolor{black}{\textbf{Bandago}} \cite{vidal2018assessing} & \textcolor{black}{Long Method} & \textcolor{black}{Java}& \textcolor{black}{4} & \textcolor{black}{10}& \textcolor{black}{Columba 1.4}& \textcolor{black}{LOC: 26,600/ 436 classes} & \textcolor{black}{\cite{Bandago-dataset}} & \textcolor{black}{Caee Study}\\
&& & & &\textcolor{black}{JGraphT 0.9.0} & LOC: 14,180 / 218 classes  & &\\
&& & & &\textcolor{black}{SportTracker 5.7} & LOC: 5,200 / 40 classes  & &\\
&& & & & \textcolor{black}{Cayanne 4.0} & LOC: 45,000 / 533 classes  & &\\
&& & & & \textcolor{black}{CheckStyle 6.4.1} & LOC: 60,000 / 399 classes  & &\\
&& & & & \textcolor{black}{Jena 2.12.1} & LOC: 54,410 / 697 classes  & &\\
&& & & & \textcolor{black}{JGroups 3.4.8} & LOC: 76,570 / 644 classes  & & \\
&& & & & \textcolor{black}{Quartz 2.1.7} & LOC: 26,810 / 176 classes  & &\\
&& & & & \textcolor{black}{Roller 5.1.2} & LOC: 47,460 / 452 classes  & &\\
&& & & & \textcolor{black}{Squirrel 3.6.0} & LOC: 79,070 / 879 classes  & & \\

   \multirow{1}{*} {\textbf{\textcolor{black}{TOAD}} \cite{antezana2019toad,alcocer2020improving}} &    Long Method &    Pharo &    N/A  &    \textcolor{black}{9} &    GitMultipileMatrix &    Unknown &    \cite{TOAD-dataset} &    Experiment\\ 
   &    &    &   &     &   TestDeviator&     &    &     \\
   &    &    &   &     &   DrTest&     &    &     \\
   &    &    &   &     &   Regis&     &    &     \\
   &    &    &   &     &   SmallSuiteGenerator&     &   &     \\
   &    &    &   &     &   Roassal&     &    &     \\
   &    &    &   &     &   Live 
 Robot Programming&     &    &     \\
   &    &    &   &     &   KerasBridge&     &   &     \\
   &    &    &   &     &   GToolkit Documenter &     &   &     \\

 \textbf{Shahidi~\etal}~\cite{shahidi2022automated} &  
 Long Method & Java & Unknown  & 5  & JEdit 4.5.1& LOC: 107,212 / 1,141 classes / 6,663 methods & Unknown & Experiment\\
& & & & & FreeMind 0.9.0 & LOC: 40,933 / 696 classes / 4,583 methods  & & \\
& & & & & ArgoUML 0.34 & LOC: 249,538 / 2,539 classes / 17,485 methods  & &\\
& & & & & JFreeChart 1.0.14 & LOC: 222,814 / 8,630 classes / 619 methods  & & \\
& & & & & jVLT 1.3.2 & LOC: 29,161 / 420 classes / 2,036 methods & &  \\

   \textbf{Segmentation}~\cite{tiwari2022identifying} &    Long Method &    Java  &     2  &    6 &    JUnit &    Unknown &    \cite{Segmentation} &       Experiment \\ 
   &   &    &    &    &   JHotDraw &     &     &     \\
   &   &    &    &    &   MyWebMarket &     &     &    \\
   &   &    &    &    &   EventBus &     &     &     \\
   &   &    &    &    &   Mockito &     &     &    \\
   &   &    &    &    &   XData &     &     &     \\

\multirow{1}{*}{\textbf{LiveRef} \cite{fernandes2022liveref,fernandes2022live}} &Long Method &Java  &  20 & 3 &Space Invaders & Unknown & \cite{LiveRef} &    Experiment \\ 
&& &&&JHotDraw & & &  \\
& & &&  &Movie rental system &  & & \\

\bottomrule
\end{tabular}
\end{adjustbox}

\vspace{-.3cm}
\end{table*}

\begin{table*}
  \centering
	 \caption{\textcolor{black}{Benchmarks and datasets used in \textit{Extract Method} refactoring studies for \textit{Code Clone} extraction.}}
	 \label{Table:Tool_Benchmark-codeclone}

\begin{adjustbox}{width=1.0\textwidth,center}
\rowcolors{2}{white}{gray!25}
\begin{tabular}{lllllllllll}\hline
\toprule
\bfseries Study & \bfseries Intent & \bfseries Language      & \bfseries No of Metric &  \bfseries No of Project & \bfseries Project & \bfseries Other Properties & \bfseries Dataset Link & \bfseries Validation Method \\
\midrule
 \textcolor{black}{\textbf{CloRT}\cite{balazinska1999partial}} & \textcolor{black}{Code Clone} & \textcolor{black}{Java} & \textcolor{black}{N/A}  & \textcolor{black}{Unknown} & \textcolor{black}{Unknown} &  \textcolor{black}{Unknown} &  \textcolor{black}{Unknown} & Proof of Concept \\ 
 
 \textcolor{black}{\textbf{Komondoor \& Horwitz} \cite{komondoor2000semantics}} &   \textcolor{black}{Code Clone} &   Procedural &  N/A  &  \textcolor{black}{Unknown} &  \textcolor{black}{Unknown} & \textcolor{black}{Unknown} &  \textcolor{black}{Unknown} &  \textcolor{black}{Proof of Concept} & \\

\textcolor{black}{\textbf{Komondoor \& Horwitz} \cite{komondoor2003effective}} &  \textcolor{black}{Code Clone} &  Procedural & N/A  & \textcolor{black}{Unknown} & \textcolor{black}{Unknown} &\textcolor{black}{Unknown} & \textcolor{black}{Unknown} & \textcolor{black}{Proof of Concept}  \\

 \textbf{CCShaper} \cite{higo2004refactoring} &  Code Clone &  Java &   6 &  1 &  Ant 1.6.0 &  LOC: 180,000 / 627 files  &  Unknown &   Caee Study\\ 

  \textbf{Aries} \cite{higo2004aries, higo2005aries,higo2008metric} & Code Clone &   Java &    6 &   1 &   Ant 1.6.0 & LOC: 180,000 / 627 files  &   Unknown &    Caee Study\\

 \textcolor{black}{\textbf{Juillerat \& Hirsbrunner}} \cite{juillerat2006algorithm} &   \textcolor{black}{Code Clone} &  Java &  N/A  &  \textcolor{black}{Unknown} &  \textcolor{black}{Unknown} &  Unknown &  \textcolor{black}{Unknown} &  \textcolor{black}{Proof of Concept}  \\

   \textcolor{black}{\textbf{Wrangler}} \cite{li2009clone} &   \textcolor{black}{Code Clone} &  Erlang/OTP  &  N/A  &  \textcolor{black}{3} &  \textcolor{black}{Wrangler} &  LOC: 30,872 &  \textcolor{black}{Unknown} &  \textcolor{black}{Experiment}  \\
    &  &  &  &   & Mnesia & LOC: 28,152 & & \\
     &  &  &  &   & Yaws & LOC: 29,603 & & \\

  \textcolor{black}{\textbf{HaRe}} \cite{brown2010clone} &   \textcolor{black}{Code Clone} &  Haskell 98  &  N/A  &  \textcolor{black}{13} &  \textcolor{black}{Previous work \cite{thompson2011haskell}} &  Unknown &  \textcolor{black}{Unknown} &  \textcolor{black}{Caee Study}  \\

\textcolor{black}{\textbf{Choi \etal}} \cite{choi2011extracting}
 & Code Clone & Java & 3 &  1 & Unknown & KLOC: 110 / 296 files & Unknown & Caee Study \\


 \textbf{CeDAR} \cite{tairas2012increasing} &  Code Clone &   Java &   2  &  9 &  Ant 1.7.0  &  KLOC: 67 &  Unknown &  Experiment  \\ 
 &  &  &  &   & Columba 1.4 &   KLOC: 75 &  &  \\
 &  &  &  &   &  EMF 2.4.1 &   KLOC: 118 & &  \\
 &  &  & &   & Hibernate 3.3.2 &  KLOC: 209 &  &  \\
 &  &  & &   & Jakarta-JMeter 2.3.2 &  KLOC: 54 &  &  \\
 &  &  & &   & JEdit 4.2 &  KLOC: 51 &  &  \\
 &  &  &  &   & JFreeChart 1.0.10 &  KLOC: 76 & &  \\
 &  &  &  &   & JRuby 1.4.0 &  KLOC: 101 & &   \\ 
 &  &  &  &   & Squirrel-SQL 3.0.3 &  KLOC: 141 & &   \\ 

 \textcolor{black}{\textbf{FTMPAT} \cite{goto2013extract}} &  \textcolor{black}{Code Clone} &  \textcolor{black}{Java} &  \textcolor{black}{3} &  \textcolor{black}{1} &  \textcolor{black}{Ant 1.7.0} &  \textcolor{black}{Unknown}  &  \textcolor{black}{Unknown} &   \textcolor{black}{Caee Study}\\

  \textbf{SPAPE} \cite{bian2013spape} &   Code Clone &   Java &     Unknown &   10 &   Linux 2.6.6/kernel &   LOC: 30,629 &   Unknown &    Experiment \\ 
  &   &   Procedural  &  &   &  Unix/make 3.82 &   LOC: 33,864  &   &  \\
  &  &   &   &   &  httpd 2.2.2/server &    LOC: 36,926  &   &  \\
  &  &   &   &   &  devecot 2.0.8/src/auth &   LOC: 18,243  &   &  \\
  &  &   &   &   &  gstreamer 0.10.31/gst &   LOC: 66,637  &   &  \\
  &  &   &   &   &  gtk 2.91.5/gdk/x11 &   LOC: 30,118  &   &  \\
  &  &   &   &   &  iptables 1.4.10/extensions &   LOC: 19,668  &   &  \\
  &  &   &   &   &  nginx-0.8.15/src/core &   LOC: 17,126  &   &  \\
  &  &   &   &   &  proftpd 1.3.3c/src &   LOC: 34,404  &   &  \\
  &  &   &   &   &  PostgreSQL 9.0.2/src/backend/access &   LOC: 65,046  &   &  \\

  \textbf{Bian} \etal \cite{bian2014identifying} &   Code Clone &   Java &     Unknown & 5 & Linux 2.6.6/arch & Unknown & Unknown & Experiment \\
 & &  &  &  &  Linux 2.6.6/net & Unknown &   &  \\
 & &  &  &  &  Linux 2.6.6/sound/drivers & Unknown &   &  \\
 & &  &  &  &  Unix/make 3.82 & Unknown &   &  \\
 & &  &  &  &  http2.2.2/server & Unknown &   &  \\
 \textbf{JDeodorant}~\cite{tsantalis2015assessing,mazinanian2016jdeodorant,tsantalis2017clone,krishnan2013refactoring,krishnan2014unification} &   \textcolor{black}{Code Clone} &  \textcolor{black}{Java} &  \textcolor{black}{N/A} &   \textcolor{black}{9} &  \textcolor{black}{Ant 1.7.0 / Ant 1.9} &  KLOC: 67 &  \textcolor{black}{Unknown} &  \textcolor{black}{Experiment}  \\ 
 & &  &  &  &  Columba 1.4 &  KLOC: 75 &   &  \\
 & &  &  &  &  EMF 2.4.1 &  KLOC: 118 &   &  \\
 & &  &  &  & JMeter 2.3.2 / JMeter 2.9 &  KLOC: 54 &   &  \\
 & &  &  &  & JEdit 4.2 &  KLOC: 51 &   &  \\
 & &  &  &  & JFreeChart 1.0.10 / JFreeChart 1.0.14 &  KLOC: 76 &   &   \\
 & &  &  &  & JRuby 1.4.0 / JRuby 1.7.3 &  KLOC: 101 &   &  \\
 & &  &  &  & Hibernate 3.3.2 &  KLOC: 209 &   &  \\
 & &  &  &  & SQuirreL SQL 3.0.3 &  KLOC: 141 &   &  \\

\multirow{1}{*} {\textbf{DCRA} \cite{arcelli2015duplicated}} & Code Clone & Java &  1 &  50 & Qualitas Corpus \cite{tempero2010qualitas} (v. 20120401) & Unknnown & Unknown&  Experiment \\ 

 \textcolor{black}{\textbf{RASE}} \cite{meng2015does} &  \textcolor{black}{Code Clone} &  \textcolor{black}{Java} &  \textcolor{black}{N/A} &  \textcolor{black}{2} &  \textcolor{black}{Previous works \cite{meng2013lase,meng2011systematic}} &  \textcolor{black}{Unknown} &  \textcolor{black}{\cite{RASE-dataset}} &  \textcolor{black}{Experiment}\\

  \multirow{1}{*}{\textbf{CREC}~\cite{yue2018automatic}} &   Code Clone &   Java &   N/A   &   6 &   Axis2 &   8,723 commits &    \cite{CREC} &   Experiment \\ 
  &   &   &  &    &  Eclipse.jdt.core&   22,358 commits &   &    \\
  &   &   &  &    &  Elastic Search&   14,766 commits &   &    \\
  &   &   &  &    &  JFreeChart&   3,603 commits &   &    \\
  &   &   &  &    &  JRuby&   24,434 commits &   &    \\
  &   &   &  &    &  Lucene &   22,061 commits &   &    \\

 \textcolor{black}{\textbf{PRI}} \cite{chen2017tool} &  \textcolor{black}{Code Clone} &  \textcolor{black}{Java} &  \textcolor{black}{N/A} & \textcolor{black}{6} &  \textcolor{black}{AlgoUML} &  \textcolor{black}{LOC: 127,145 / 1,559 files} &  \textcolor{black}{Unknown}  &  \textcolor{black}{Caee Study} \\
 & &  &  &  & Tomcat &  LOC: 215,584 / 1,537 files &   &  \\
 & &  &  &  & Log4j &  LOC: 59,499 / 817 files &   &  \\
 & &  &  &  & Eclipse AspectJ &  LOC: 326,563 / 4,758 files &   &  \\
 & &  &  &  & JEdit &  LOC: 107,368 / 561 files &   &  \\
 & &  &  &  & JRuby &  LOC: 186,514 / 1,256 files &   &  \\

\textcolor{black}{\textbf{Ettinger \etal}} \cite{ettinger2016duplication,ettinger2017efficient} &  \textcolor{black}{Code Clone} & Java & N/A & Unknown & Previous work \cite{tiarks2011extended} & 59 clone pairs &Unknown & Proof of Concept \\

 \multirow{1}{*} {\textbf{Unnamed}~\cite{yoshida2019proactive}} &  Code Clone  &  Java   &  N/A   &  2 &  JFreeChart &  KLOC: 260 / 990 classes  &  Unknown &  Experiment \\ 
 & &  &  &  & JUnit &  KLOC: 43 / 449 classes &   &   \\

  \textcolor{black}{\textbf{Unnamed}} \cite{shin2019study} &   \textcolor{black}{Code Clone} &   \textcolor{black}{Java}&   \textcolor{black}{N/A} &   \textcolor{black}{Unknown}&   \textcolor{black}{Unknown}&   \textcolor{black}{Unknown} &   \textcolor{black}{Unknown} &   \textcolor{black}{Caee Study}\\

  \textbf{CloneRefactor} \cite{baars2019towards} & Code Clone &  Java & N/A & 1.343 & Previous work \cite{allamanis2013mining} & LOC (AVG): 980 & Unknown & Experiment\\

  \textbf{Sheneamer}  \cite{sheneamer2020automatic} &  Code Clone
&   Java &  N/A &  6 &  Previous work \cite{yue2018automatic} &  \cite{CREC} &  Dataset of \cite{CREC} &  Experiment \\
 &  &  &  &  6 &  netbeans &  200 paired clones &   Unknown &    \\
 &  &  &  &  &  eclipse-jdtcore &  400 paired clones &   &   \\
 &  &  &  &  &  EITC &  426 paired clones &   &  \\
 &  &  &  &  &  J2sdk1.4.0-javax &  482 paired clones &   &  \\
 &  &  &  &  &  eclipse-ant &  522 paired clones &   &  \\
 &  &  &  &  &  cocoon &  655 paired clones &   &  \\

  \textbf{AntiCopyPaster} \cite{alomar2022anticopypaster,alomar2023just} &   Code Clone &   Java &    78  &   13 &   arthas  &   73,884 total commits &  \cite{AntiCopyPaster-dataset} &   Experiment \\
  &  &   &   &   &  easyexcel &   &  &   \\
  &  &   &   &    &  camel-quarkus &   &  &   \\
  &  &   &   &    &  commons-lang &   &  &   \\
  &  &   &   &   &  flink &   &  &   \\
  &  &   &   &    &  iceberg &   &  &   \\
  &  &   &   &   &  jena &   &  &   \\
  &  &   &   &  &   pulsar &   &  &   \\
  &  &   &   &    &  storm &   &  &   \\
  &  &   &   &    &  apollo &   &  &   \\
  &   &   &  &    &  JavaGuide &   &  &    \\

\bottomrule
\end{tabular}
\end{adjustbox}

\vspace{-.3cm}
\end{table*}


\begin{table*}
  \centering
	 \caption{\textcolor{black}{Benchmarks and datasets used in \textit{Extract Method} refactoring studies for \textit{Separation of Concerns}.}}
	 \label{Table:Tool_Benchmark-SoC}

\begin{adjustbox}{width=1.0\textwidth,center}
\rowcolors{2}{white}{gray!25}
\begin{tabular}{lllllllllll}\hline
\toprule
\bfseries Study & \bfseries Intent & \bfseries Language      & \bfseries No of Metric &  \bfseries No of Project & \bfseries Project & \bfseries Other Properties & \bfseries Dataset Link & \bfseries Validation Method \\
\midrule
\textbf{Maruyama}~\cite{maruyama2001automated} & 
Separation of Concerns 
&  Java & N/A & \textcolor{black}{Unknown} & \textcolor{black}{Unknown} &\textcolor{black}{Unknown} & \textcolor{black}{Unknown} & \textcolor{black}{Proof of Concept}   \\
 \textcolor{black}{\textbf{Nate}} \cite{ettinger2004untangling} &  \textcolor{black}{Separation of Concerns} &  \textcolor{black}{Java} &  \textcolor{black}{N/A} &  \textcolor{black}{Unknown} &  \textcolor{black}{Unknown} &  \textcolor{black}{Unknown} &  \textcolor{black}{Unknown} &  \textcolor{black}{Proof of Concept} \\

\textbf{SDAR} \cite{o2005star} & Separation of Concerns & Java & N/A  & Unknown & Unknown &Unknown & Unknown&  Proof of Concept \\ 

\textcolor{black}{\textbf{Juillerat \& Hirsbrunner}} \cite{juillerat2007improving} &  \textcolor{black}{Code Clone} & Java & N/A  & \textcolor{black}{Unknown} & \textcolor{black}{Unknown} & Unknown & \textcolor{black}{Unknown} & \textcolor{black}{Proof of Concept}  \\

 \textbf{Xrefactory} \cite{vittek2007c++} &  Separation of Concerns &  C++ &   N/A &  Unknown&  Unknown &   Unknown &  Unknown &  Proof of Concept \\ 


\textbf{Unnamed} \cite{corbat2007ruby} & Separation of Concerns & Ruby & N/A & Unknown & Unknown &  Unknown & Unknown & Proof of Concept \\ 

 \textbf{RefactoringAnnotation} \cite{murphy2008breaking} &  Separation of Concerns & Java &  Unknown  &  5 &  Azureus &  Unknown & Unknown &   Experiment \\ 
 & &  &  &  &  GanttProject &   &   &   \\
 & &  &  &  &  JasperReports &   &   &   \\
 & &  &  &  &   &   &   &   \\
 & &  &  &  &  Java 1.4.2 libraries &   &   &   \\

\textbf{Abadi \etal} \cite{abadi2008re} &
Separation of Concerns
&  \textcolor{black}{Java} & N/A & \textcolor{black}{Unknown} & \textcolor{black}{Unknown} & Unknown & \textcolor{black}{Unknown} & \textcolor{black}{Caee Study} \\

\textbf{Abadi \etal} \cite{abadi2009fine} &
Separation of Concerns
&  \textcolor{black}{Java} & N/A & \textcolor{black}{Unknown} & \textcolor{black}{Unknown} & Unknown & \textcolor{black}{Unknown} & \textcolor{black}{Caee Study} \\

\multirow{1}{*} {\textbf{ReAF}~\cite{kanemitsu2011visualization}} &  Separation of Concerns & Java   &   Unknown &  1 &  Ant 1.8.1 & Unknown & Unknown & Experiment  \\ 

 \textbf{Sharma}~\cite{sharma2012identifying} & 
 Separation of Concerns & C/C++ & N/A & 1 & CppCheck & Unknown & Unknown & Proof of Concept \\

 \textbf{Unnamed} \cite{cousot2012abstract} &  Separation of Concerns &  C\# &  Unknown &  Unknown &   Unknown &  Unknown &  Unknown&   Proof of Concept \\ 

  \textbf{JExtract}~\cite{silva2014recommending,silva2015jextract} &   Separation of Concerns &   Java &    Unknown 
 &   12 &   MyWebMarket &  Unknown &   \cite{JExtract} &   Experiment \\ 
   &   &   &  &    &  JUnit 3.8 / 4.10&   &   &    \\ 
  &   &   &  &    &  JHotDraw 5.2 &   &   &    \\ 
  &   &   &  &    &  Ant 1.8.2 &   &   &    \\ 
  &   &   &  &    &  ArgoUML 0.34 &   &   &    \\ 
  &   &   &  &    &  Checkstyle 5.6 &   &   &    \\ 
  &   &   &  &    &  FindBugs 1.3.9 &   &   &    \\ 
  &   &   &  &    &  FreeMind 0.9.0 &   &   &    \\ 
  &   &   &  &    &  JFreeChart 1.0.13&   &   &    \\ 
  &   &   &  &    &  Quartz 1.8.3 &   &   &    \\ 
  &   &   &  &    &  SQuirreL SQL 3.1.2 &   &   &    \\ 
  &   &   &  &    &  Tomcat 7.0.2 &   &   &    \\

 \textbf{GEMS}~\cite{xu2017gems} &  Separation of Concerns &  Java &    48  & 5 &  Wikidev &  56 methods & Unknown &  Experiment\\ 
 &  &  & &   & SelfPlanner& 25 methods &  &   \\
 &  &  & &   & MyWebMarket&   23 methods &  &   \\
 &  &  & &   & JUnit&  12 methods &  &   \\
 &  &  & &   & JHotDraw &  14 methods & &   \\

 \textbf{Imazato \etal} \cite{imazato2017finding} &   Separation of Concerns &  Java & & 5 & Ant & LOC: 260,624 / 1,532 methods & Unknown  & Experiment  \\
& & & & &  ArgoUML & LOC: 370,750 / 1,470  methods & &\\
& & & & &  JEdit & LOC: 187,166 / 1,066 methods & &\\
& & & & &  jFreeChart & LOC: 327,865 / 180 methods & &\\
& & & & &  Mylyn & LOC: 166,149 / 980 methods & &  \\

  \textbf{PostponableRefactoring} \cite{maruyama2017tool} & Separation of Concerns &   Java &   N/A &   Unknown &   Unknown &    Unknown &   Unknown &   Proof of Concept \\ 

\textbf{Nyamawe \etal} \cite{nyamawe2019automated,nyamawe2020feature} 
& Separation of Concerns
&  Java & N/A & 55 & \cite{Nyamawe} & Unknown & \cite{Nyamawe} & Experiment    \\

 \textbf{Krasniqi \& Cleland-Huang} \cite{krasniqi2020enhancing} &   
 Separation of Concerns & Java & N/A & 4 & Derby & KLOC: 170/ 2,382 commits & \cite{Krasniqi} & Experiment \\
& & & & & Drools & KLOC: 371 / 840 commits & &\\
& & & & & Groovy & KLOC: 141 / 4,892 commits & & \\
& & & & &  Infinispan & KLOC: 299 / 2,349 commits & & \\

\textbf{Abid \etal} \cite{abid2020does} 
& Separation of Concerns
& Java & 8 & 30 &  \cite{Chima} & Unknown & \cite{Chima} & Experiment\\

\textbf{Aniche~\etal}~\cite{aniche2020effectiveness}
& Separation of Concerns
&  Java & 61 & 11,149  & \cite{Aniche} & 8.8
million commits & \cite{Aniche} & Experiment \\

 \textbf{Van~der~Leij~\etal}~\cite{van2021data} &   
 Separation of Concerns &
Java & 7 & 11,149 & Previous work \cite{aniche2020effectiveness} & 8.8
million commits & Dataset of \cite{aniche2020effectiveness} & Experiment  \\

\textbf{Sagar \etal} \cite{sagar2021comparing} & 
Separation of Concerns & Java & 60 & 800 & Previous work \cite{alomar2021we} &  748,001 commits & Dataset of \cite{alomar2022documentation} & Experiment \\

 \textbf{AlOmar \etal} \cite{alomar2022documentation} &  
 Separation of Concerns & Java & N/A & 800 & Previous work \cite{alomar2021we} & 748,001 commits & \cite{SAR2020WEB} & Experiment \\
  \\

\textbf{Nyamawe} \cite{nyamawe2022mining}  
& Separation of Concerns
& Java & N/A & 65 & Previous works \cite{krasniqi2020enhancing,nyamawe2020feature,rebai2020recommending} & 7,520 commits & Datasets of \cite{krasniqi2020enhancing,nyamawe2020feature,rebai2020recommending}  & Experiment \\

 \textbf{Cui \etal} \cite{cui2023rems} & 
 Separation of Concerns & Java & N/A & Unknown & Previous works \cite{Silva2016why,xu2017gems}& Unknown &\cite{REMS} &  Experiment  \\
 
\textcolor{black}{\textbf{REM} \cite{thy15adventure}} & \textcolor{black}{Separation of Concerns} & \textcolor{black}{Rust} & \textcolor{black}{N/A} & \textcolor{black}{5} &   \textcolor{black}{petgraph} &  \textcolor{black}{LOC: 20,157} & \textcolor{black}{\cite{REM}} & \textcolor{black}{Caee Study}  \\
& & & & & \textcolor{black}{gitoxide} & LOC: 20,211 & &\\
& & & & & \textcolor{black}{kickof} & LOC: 1,502 & &\\
& & & & & \textcolor{black}{sniffnet} & LOC: 7,304 & & \\
& & & & & \textcolor{black}{beerus} & LOC: 302 & & \\ 

\textbf{Palit \etal} \cite{palit14automatic} & Separation of Concerns & Java & 61 & 410 & Previous work \cite{aniche2020effectiveness} & 55,268 commits & \cite{Palit} & Experiment\\
\bottomrule
\end{tabular}
\end{adjustbox}

\vspace{-.3cm}
\end{table*}

\begin{boxK}
\textit{\textbf{Summary.}  \textcolor{black}{Out of the 83 primary studies analyzed, almost 78\% of the datasets are  not publicly available. There is a lack of sharing datasets, which is detrimental to reproducing research.  Primary studies have mostly employed small or medium-scale open-source applications, often developed using Java, typically containing less than 225,000 lines of code. These datasets are heterogeneous and do not contain the same type of information, making their standardization, for the purpose of benchmarking, difficult.}
} 
\end{boxK}



\section{Discussion and Open Issues}
\label{Section:Reflection}

To ensure that the \textit{Extract Method} refactoring is properly identified/applied, we recommend \textcolor{black}{retrofitting} these tools with the following dimensions:

 \faThumbTack \noindent{\textbf{ Provide context to guide developers on how to use \textit{Extract Method} refactoring tools.}}
\textcolor{black}{Based on the findings from RQ$_1$ and RQ$_2$, it becomes apparent that certain tools offer the context in which the  \textit{Extract Method} refactoring is being performed} (\eg \texttt{JDeodorant}, \texttt{SEMI}, \texttt{AntiCopyPaster}). The opportunities of applying this refactoring might be related to \textit{Duplicate Code} removal, \textit{Long Method} extraction, etc. However, other tools (\eg \texttt{ReAF}, \texttt{SDAR}) lack the context in which the \textit{Extract Method} is being performed. It is worth noting that without properly considering the context, the ground truth used to compare against existing studies might be biased. Also, the comparison against the state-of-the-art may not be appropriate unless these tools are called in the same context or intent as their original papers. For instance, \texttt{JDeodorant} applies the \textit{Extract Method} refactoring to deal with long methods. If this tool is being tested against an \textit{Extract Method} performed to remove duplicates, it is expected not to recommend any code changes. Therefore, performing experimentation against techniques tackling different intents may not be adequate. In a similar context, building a universal model that extracts methods based on the history of code changes without understanding the intent must be human-verified to see whether it is useful. 

\faThumbTack \noindent{\textbf{ Recommend appropriate naming for the method after the extraction.}} Since the main purpose of the tools listed in Table \ref{Table:Tool_Characteristics} is the recommendation of \textit{Extract Method} refactoring, developers will ultimately need to provide a clear name for the extracted method, which is considered one of the most influential factors in the developer’s decision on whether to perform \textit{Extract Method} or not \cite{yamanaka2021recommending,bavota2014recommending}. The appropriate name assists in expressing its role and meaning to the extracted code. The existing approaches can complement their recommendation of the \textit{Extract Method} with the naming recommendation of the extracted method. 

\faThumbTack \noindent{\textbf{ Lack of clarity of how the approaches leverage metrics and decide the associated threshold to make the decision.}} From Figure \ref{fig:metric}, we observe different software quality metrics related to various quality attributes used by the tools. For instance, \texttt{AntiCopyPaster} has used 78 metrics related to size, complexity, coupling, and keywords to extract duplicate code. In contrast, \texttt{LiveRef} utilized around 20 metrics related to complexity, cohesion, and maintainability to identify the extraction targets of \textit{Long Method} code smell. However, the implementation of these metrics
may vary between these tools based on the context. In addition, there may be cases where different metric names are used to improve some quality attributes. This phenomenon might impact the interpretation of the correctness of the recommended tools.

\faThumbTack \noindent{\textbf{ Adapt \textit{Extract Method} refactoring operations for multiple programming languages.}} As reported in 
RQ$_2$,  \textcolor{black}{there are} an existence of multiple \textit{Extract Method} refactoring tools; however, 
RQ$_1$ and RQ$_2$ findings show that most of these tools are limited to supporting Java systems which
narrow \textit{Extract Method}-related research to Java systems. Hence, restricting
research to a single language will not accurately reflect real-world scenarios \cite{silva2020refdiff}; there are opportunities for researchers to evolve the field further and increase the diversity of their research. The developers of non-Java systems gain no benefit without a tool to use in their development workflow. Furthermore, recent trends
have shown a rise in the popularity of dynamically typed programming languages (\eg Python), giving more urgency for the research community to construct tools that support non-traditional research languages.



\faThumbTack \noindent{\textbf{ Lack of benchmarks.} With the rise of refactoring mining tools \cite{prete2010template,tsantalis2018accurate,silva2020refdiff}, such tools were used to create datasets that already performed \textit{Extract Method} refactorings from open-source software repositories. The collected refactorings became one of the main sources \textcolor{black}{of already} 
 quantitative analysis for refactoring recommendation studies. For instance, the mined \textit{Extract Method} refactorings were used either as an oracle to validate the correctness of recommendations \cite{mkaouer2014recommendation,mkaouer2015many,silva2014recommending,silva2015jextract}, or as training and testing sets for machine learning models and deep learning models \cite{nyamawe2022mining,alomar2022anticopypaster,alomar2023just}. While these tools have demonstrated high detection accuracy \cite{tsantalis2020refactoringminer}, they solely parse source code changes to identify refactoring patterns. So, there is no association between the performed refactoring and the developer's rationale behind it. Even the reliance on the developer's documentation of the code change may not necessarily reveal the needed details behind the refactoring intent. Without such information, it becomes difficult to guess whether a mined \textit{Extract Method} was performed to split a long method, segregate concerns from a complex method, or remove a clone. Therefore, studies using these data sets make assumptions concerning their intent, which may or may not hold. Any refactoring being performed outside of the paper's presumed context is noise that may hinder the data quality for training or validation. That is why it is essential to curate any collected refactorings by associating them with their proper context. Yet, the task of labeling refactorings' contexts may not be trivial.}

\faThumbTack \noindent{\textbf{ Lack of clarity on potential \textit{Extract Method} drawbacks.}} All reviewed studies primarily focus on motivating the need for method extaction to improve readability, maintainability, and reusability. However, it is critical to raise the developer's awareness of the potential limitations inherited from the solutions' design or execution. One of the main design-level limitations of these approaches is the potential increase in the code's cognitive complexity. In fact, when a new method is extracted, it may introduce additional local variables and parameters. Such addition can adversarially hinder program comprehension and add a maintenance burden. Additionally, adding new method calls comes with additional overhead, such as method dispatch and return, which may reduce the program's performance, especially when the extracted code breaks tight loops \cite{huang2014performance}. Finally, depending on where the extracted method lives, it can change the scope or visibility of its variables or objects, leading to a violation of the behavior preservation property. While the benefits of the proposed refactorings may outweigh the drawbacks, studies should warn developers to avoid introducing regressions in their systems.

\faThumbTack \noindent{\textbf{ Integration of \textit{Extract Method} tools into the developer workflow.}} While our finding from RQ$_2$ shows that researchers proposed an approach to recommend \textit{Extract Method} refactoring opportunities, not all approaches can be used in practice. Hence, the community needs to better collaborate with established tool/IDEs vendors in integrating their contributions with popular tools and IDEs
to promote the usage of their artifacts. As for the existing tools, in addition to providing extensive and innovative refactoring functionality, researchers must ensure that their products exhibit an optimal user experience. Usability and trustworthiness are essential to refactoring tool adoption and are among the reasons for the \textcolor{black}{limited usage} \cite{eilertsen2021usability,murphy2011we, vakilian2014alternate, vakilian2012use}.

\faThumbTack \noindent{\textbf{\textbf{ \textcolor{black}{\textit{Extract Method} refactoring support using Large Language Models (LLMs).}}} 
\textcolor{black}{While \textit{Extract Method} is considered as one of the most popular refactoring operations and represents approximately 49.6\% of the total refactorings recommended \cite{Jdeodorant}, it is recognized as one of the most difficult and error-prone refactorings \cite{Silva2016why,murphy2011we,golubev2021one}. Even though we have shown in this systematic review multiple studies on \textit{Extract Method} in the literature using multiple artificial intelligence (AI) techniques, its adoption is still challenging for developers \cite{Silva2016why,murphy2011we}. 
More recently, Large Language Models (LLMs) have made rapid advancements that have brought AI to a new level, enabling and empowering even more diverse software engineering applications and industrial domains with intelligence \cite{fan2023large, vaithilingam2022expectation,xia2023automated,zhang2023sentiment,white2023chatgpt}. Such LLMs are pre-trained on large corpora of data which enclose numerous commonsense knowledge and support Transformer architecture with millions, even billions of parameters. We believe that the \textit{Extract Method} can benefit significantly from LLM advances. For instance, dedicated LLMs can be used to identify code fragments that need to be extracted and to recommend appropriate names for the extracted methods. LLMs can also automatically generate the documentation of \textit{Extract Method} refactoring changes, \eg generate the commit message or pull request description along with the intent behind the refactoring. It can also help with code review by explaining the intent of the \textit{Extract Method} refactoring and providing a summary of the code change before and after the refactoring. We thus believe that LLMs represent a unique technique to empower \textit{Extract Method} refactoring and open up various research venues in the field of \textit{Extract Method} in particular and refactoring in general.}

\section{Threats to Validity}
\label{Section:Threats}

In this section, threats are discussed in the context of three types of threats of validity: internal validity, construct validity, and external validity.


\noindent{\textbf{Internal threats to validity:}}
\textcolor{black}{Obtaining a representative set of literature publications for this SLR can be considered a validity threat due to the search process. To minimize this threat, we followed the SLR guidelines  \cite{Kitchenham07guidelinesfor,wohlin2014guidelines,kitchenham2004procedures,brereton2007lessons,kitchenham2009systematic}. In particular, we have carefully established search engines, search terms, and inclusion/exclusion criteria to ensure that the review of the literature is comprehensive. Additionally, we considered related search terms and the main terms of the research questions to construct the search string and select relevant articles. Furthermore, we followed a five-stage study selection process and applied each stage's inclusion and exclusion criteria described in Section \ref{Section:methodology}. Moreover, the analysis involved snowballing to expand the paper collection. These study design steps reduce the possibility that papers are missed. Another threat is the limitation of search terms and search engines, which might lead to incomplete literature publications. To limit this threat, we used carefully defined keywords and comprehensive academic search engines (\ie ScienceDirect, Scopus, Springer, Web of Science, ACM, IEEE, and Wiley) that cover the main publishers' venues. We observed that when using search engines, particularly IEEE, some papers containing our keywords were not being found despite being indexed in their libraries. This issue has been reported in previous studies when using the IEEE search engine \cite{landman2017challenges,zakeri2023systematic}. However, we found these missed papers during the snowballing process. Regarding the quality of the selected PSs, only the studies that underwent peer review by leading academic publishers were included. Furthermore, selected studies that were within the search timeline were included. To our knowledge, all PSs relevant to our research goal and within the search window have been included.}

\noindent{\textbf{Construct threats to  validity:}}
\textcolor{black}{Concerning the subjectivity of the assessment of the PSs, the primary studies were reviewed independently by
two authors. The first author performed data analysis and extraction from the second author, who reviewed the currently selected PSs. At the end of each iteration, the authors met and performed any necessary refinements. In the event of disagreements, the researchers discussed these cases to reach a consensus. Furthermore, to avoid personal bias during manual analysis, two authors conducted each step in the manual analysis, and the results were always cross-validated.} \textcolor{black}{Moreover, some PSs do not make a clear distinction between how 
 refactoring opportunities are detected, and how the refactoring is actually performed. Therefore, for these studies, we consider detection to refactoring opportunities to be part of the correction if the end goal of the PSs is \textit{Extract Method} refactoring identification.}

\noindent{\textbf{External threats to  validity:}}
\textcolor{black}{The collected papers contain a significant proportion of academic works, forming an adequate basis for concluding findings that could be useful for academia. However, we cannot claim that the same \textit{Extract Method} detection and execution is used in industry. Additionally, our findings are mainly within the field of software refactoring. We cannot generalize our results beyond this subject.}

\section{Conclusion}
\label{Section:Conclusion}
In this paper, we map and review the body of knowledge on \textit{Extract Method} refactoring opportunities.  We systematically reviewed \finalpool papers and classified them. This research aims to aggregate, summarize and discuss the practical approaches that recommend \textit{Extract Method} refactoring. Our main findings show that \textcolor{black}{(i) 38.6\% of \textit{Extract Method} refactoring studies
primarily focus on addressing code clones;
(ii) Several of the \textit{Extract Method} tools involve the developer in the decision-making process when applying the method extraction, and (iii)  the existing benchmarks vary widely and lack uniform information, posing challenges in standardizing them for benchmarking purposes.} This existing research empowers the community with information to guide future \textit{Extract Method} tool development. Future work includes
evaluation of each tool to determine the extent to which tools recommend \textit{Extract Method} refactoring given the same context.


\section*{Acknowledgments}
\label{Section:Acknowledgments}

The authors sincerely thank the anonymous reviewers for their invaluable feedback and constructive comments, which enhanced the quality and rigor of this work. Their thoughtful insights and suggestions have been instrumental in shaping the final version of this paper.

This research is partially by the National Science Foundation under Grant No. CNS-2213765.

\bibliographystyle{ieeetr}
{\scriptsize\bibliography{references}}

\end{document}